\documentclass{aa}

\usepackage{fabien}
\usepackage[breaklinks,colorlinks,citecolor=blue]{hyperref}

\usepackage{natbib}
\bibpunct{(}{)}{;}{a}{}{,}

\begin{document}

\title{Fast and easy super-sample covariance of large scale structure observables}
\titlerunning{Fast and easy SSC}

\author{Fabien Lacasa\thanks{fabien.lacasa@unige.ch}\inst{\ref{inst1}} \and Julien Grain\inst{\ref{inst2}}}
\institute{
D\'{e}partement de Physique Th\'{e}orique and Center for Astroparticle Physics, Universit\'{e} de Gen\`{e}ve, 24 quai Ernest Ansermet, CH-1211 Geneva, Switzerland\label{inst1} \and Institut d'Astrophysique Spatiale, CNRS (UMR8617) and Universit\'{e} Paris-Sud 11, B\^{a}timent 121, 91405 Orsay, France\label{inst2}
}

\date{\today}

\abstract
{
We present a numerically cheap approximation to super-sample covariance (SSC) of large scale structure cosmological probes, first in the case of angular power spectra. It necessitates no new elements besides those used for the prediction of the considered probes, thus relieving analysis pipelines from having to develop a full SSC modeling, and reducing the computational load. The approximation is asymptotically exact for fine redshift bins $\Delta z \rightarrow 0$. We furthermore show how it can be implemented at the level of a Gaussian likelihood or a Fisher matrix forecast, as a fast correction to the Gaussian case without needing to build large covariance matrices. Numerical application to a Euclid-like survey show that, compared to a full SSC computation, the approximation recovers nicely the signal-to-noise ratio as well as Fisher forecasts on cosmological parameters of the $w$CDM cosmological model. Moreover it allows for a fast prediction of which parameters are going to be the most affected by SSC and at which level. In the case of photometric galaxy clustering with Euclid-like specifications, we find that $\sigma_8$, $n_s$ and the dark energy equation of state $w$ are particularly heavily affected. We finally show how to generalize the approximation for probes other than angular spectra (correlation functions, number counts and bispectra), and at the likelihood level, allowing for the latter to be non-Gaussian if needs be. We release publicly a Python module allowing to implement the SSC approximation, as well as a notebook reproducing the plots of the article, at \url{https://github.com/fabienlacasa/PySSC}
}

\maketitle


\section{Introduction}\label{Sect:intro}

The matter distribution at large scales in the Universe is one of the main cosmological probes allowing for shading lights on e.g. the dark matter, dark energy, and gravity at cosmological scales. The current surveys of galaxies such as KiDS \citep{Hildebrandt2017} and the Dark Energy Survey (DES) \citep{DES2017,Troxel:2017xyo} recently provided cosmological constraints on the based $\Lambda$CDM model from galaxy clustering and weak lensing which are now competitive with constraints derived from the lensing of the Cosmic Microwave Background (CMB) and consistent with CMB primary anisotropies \citep[for a recent comparison, see e.g.][]{Aghanim:2018eyx}. In the near future, large surveys such as the Large Synoptic Sky Telescope \citep[LSST,][]{Abell:2009aa} and the Euclid satellite mission \citep{Euclid-redbook} will greatly improve our understanding of the structuration of the Universe, the nature and the properties of dark energy, potential modification of gravity at cosmological scales, and the initial conditions of cosmological perturbations \citep{Amendola:2012ys}.

Unlike CMB primary anisotropies however, late-time tracers of the large scale structures (LSS) evolved through non-linear dynamics, and as a result, the probability distribution function of probes such as the galaxy distribution or weak lensing by LSS is no more Gaussian, with deviation from a Gaussian distribution increasing at smaller scales. This first means that not all the information is compressed in the two-point statistics of the considered probes. Second, this means that the covariance of statistical observables built form LSS tracers (e.g. any $n$-point statistics) is increased by the presence of non-Gaussian contributions. (As an example, the covariance on angular power spectra will be increased by contributions from a non vanishing trispectrum.) In the present context of preparing the cosmological interpretation of forthcoming datasets, as well as forecasting the expected performances of future galaxy surveys which aims at precision cosmology from LSS tracers, it is now necessary to properly take into account the non-Gaussian contribution to the covariance for any inference of cosmological parameters from LSS observables. \\

Among the different non-Gaussian sources to the covariance (see \citet{Lacasa2018b} for a full derivation of them) is the super-sample covariance (SSC), first discovered for cluster counts by \citet{Hu2003} and to which a vast literature has been devoted \citep[e.g.][]{Takada2013,Takada2014,Takahashi2014,Li2017,Chan2018,Lacasa2018,Barreira2018,Barreira2018b}. This additional source of cosmic variance is inherent to all galaxy surveys due to the limited portion of the Universe which is observed, both in redshift depth and in sky fraction. SSC hence comes from the non-linear impact of density fluctuations with wavelengths greater than the survey size. These super survey modes modulate the local observables by making the background density averaged over the survey size to be non-representative (either denser or less dense than) of the averaged density in the Universe. Barring systematics, SSC is expected to be the dominant source of statistical error / cosmic variance for weak lensing \citep{Barreira2018b} beyond the usual Gaussian covariance, although other terms may also be important for galaxy clustering \citep{Lacasa2018b}. It affects the whole set of statistical observables and correlates them. Contrary to intrasurvey sources of covariance, it can be shown that SSC cannot be reliably calibrated from data itself nor from classical simulations \citep{Lacasa2017}. This thus motivates the need for analytical or semi-analytical predictions of the effect, for use in the analysis of current and future galaxy surveys.

When analyzing such galaxy surveys, we usually deal with observables $\mathcal{O}_i$ being line-of-sight integrals of the form
$\mathcal{O}_i = \int \dd V_{i} \, \mathfrak{o}_i$,
where $\mathfrak{o}_i$ is the comoving density of the observable (including selection effects such as redshift binning), and $\dd V = r^2(z) \frac{\dd r}{\dd z} \dd z$ is the comoving volume per steradian. Then the rigorous super-sample covariance for such observables is given by \citep[e.g.][]{Lacasa2016}
\ba \label{Eq:exact-SSC}
\Cov_\mr{SSC}\left(\mathcal{O}_1,\mathcal{O}_2\right) = \iint \dd V_{1} \dd V_2 \, \frac{\partial \mathfrak{o}_1}{\partial\delta_b}(z_1) \, \frac{\partial \mathfrak{o}_2}{\partial \delta_b}(z_2) \, \sigma^2(z_1,z_2).
\ea
In the above, $\frac{\partial \mathfrak{o}_1}{\partial\delta_b}(z_1)$ is the {\it response} of the probe which amounts how a given probe varies with changes of the background density $\delta_b$. The quantity $\sigma^2(z_1,z_2)$ reads (assuming full sky here for simplicity)
\ba \label{Eq:sigmasquare}
\sigma^2(z_1,z_2) = \frac{1}{2\pi^2}\int k^2 \, \dd k \ P_\mr{m}(k|z_{12}) \ j_0(k r_1) j_0(k r_2),
\ea
with $P_\mr{m}(k|z_{12})$ the linear matter cross-spectrum between redshifts $z_1$ and $z_2$, and $j_0$ the spherical Bessel functions. It basically amounts the variation of background density on a given survey volume due to super survey modes modulations.\\
Computing exactly this SSC contribution to the covariance becomes however rapidly costly. In practice, one takes advantage of the separability in redshift \citep[e.g.][]{Lacasa2018,Barreira2018} to reduce the cost of a covariance evaluation to that of an angular power spectrum evaluation. However the covariance needs to be evaluated at every pair of multipoles. For future surveys doing angular power spectra analysis with $\ell_\mr{max}$ of a few thousands, this induces a $\mathcal{O}(10^3)$ slow-down of prediction pipelines, which can be increased by more orders of magnitude if we include tomography ($\Rightarrow$ pairs of redshift bins) and combine probes ($\Rightarrow$ pairs of probes).\\
Furthermore an exact computation necessitates the knowledge of the probe's response $\frac{\partial \mathfrak{o}}{\partial \delta_b}$, either through analytical means or through simulations, for every redshift and multipoles, which is a barrier for analysts not already experts in the field of SSC. It is thus desirable to have instead simpler functions, if not fixed parameters as we will find later on, with reference ansatzs that can be easily implemented by the community.

The aim of this article is thus to present an approximation for the SSC allowing for fast numerical computation and ease of use by the community, and to assess its accuracy in a forecast analysis using the Fisher matrix approach. \\

The article is organized as follows. Our approximation is presented in Sect. \ref{Sect:fast-approx} for the case of angular power spectra as our statistical observables. This approximation basically abolishes the above-mentioned numerical burden, and makes the computation of the super-sample covariance matrix as fast as the computation of the involved angular power spectra. Furthermore, we will show in Sect. \ref{Sect:applic} that the resulting matrix form enables fast application to common uses of the covariance (i.e. in a Gaussian likelihood or for computation of a signal-to-noise ratio or a Fisher matrix), as a correction to the Gaussian case. Then in Sect. \ref{Sect:Num-Fisher} we will show numerical results validating the approximation and giving its range of applicability. Finally in Sect. \ref{Sect:generalisations} we will generalize the approach to other statistics (number counts, correlation function and bispectrum) and to the full likelihood, making the implementation of super-sample covariance feasible even if the likelihood is not Gaussian.

We release publicly a Python code that allows to easily implement SSC with our approach at \url{https://github.com/fabienlacasa/PySSC}


\section{Approximating  the SSC}\label{Sect:fast-approx}

We consider the case of the angular power spectra cross-correlating two LSS tracers, labelled $A$ and $B$. In the context of galaxy surveys, these two tracers typically are  galaxy clustering and galaxy shear. This can however be extended to other LSS tracers such as lensing of the CMB or the integrated Sachs-Wolfe effect (iSW). This signals are observed in some redshift bins labelled by indices $i_z,\,j_z,$ etc. and with a given width. In full generality, the redshift bins may overlap.\footnote{This is for instance the case for the shear signals and the iSW since they are integrated signals from the redshift of the source plane to the observer.}

We use the Limber approximation throughout the article, both for the power spectrum and the super-sample covariance. The approximation is accurate enough for the power spectrum on the range of scales of our later forecast ($\ell\geq50$). Furthermore, it is even more adapted to super-sample covariance, because SSC impacts the covariance on small scales $\ell\gtrsim300$ as we will see in Sect.~\ref{Sect:Num-Fisher}.

With Limber approximation, the angular power spectrum between two signals can generally be written as 
\ba \label{Eq:aps}
C_{\ell}^{AB}(i_z,j_z) = \int \dd V \ W^A_{i_z}(z) \, W^B_{j_z}(z) \ P_{AB}(k_\ell|z).
\ea
The weighting kernels $W^A_{i_z}(z)$, $W^B_{j_z}(z)$ are nonzero over the width of the redshift bin, and they have unit : [probe unit] $\cdot$ sr/(Mpc/h)$^3$. The quantity $P_{AB}(k_\ell|z)$ is the 3D power spectrum of the considered probe, evaluated at the Limber wavenumber $k_\ell = (\ell+1/2)/r(z)$ with $r(z)$ the comoving distance. Weighting kernels and power spectra for the different probes of interest (galaxy clustering and shear, CMB lensing, iSW effect) are given in App. \ref{App:weighting-kernels}. \\

For an angular power spectrum, the comoving density of the observable, i.e. $\mathfrak{o}_{AB}$ entering Eq. (\ref{Eq:exact-SSC}), is  $\mathfrak{o}_{AB} = W^A_{i_z}(z) \, W^B_{i_z}(z) \ P_{AB}(k_\ell|z)$. Assuming that in Eq. (\ref{Eq:exact-SSC}) the responses, $\frac{\partial \mathfrak{o}_{AB}}{\partial\delta_b}$, vary slowly with redshift compared to $\sigma^2(z_1,z_2)$, we arrive at the approximation at the base of this article:
\ba
\nonumber \Cov_\mr{SSC}\left(C_{\ell}^{AB}(i_z,j_z),C_{\ell'}^{CD}(k_z,l_z)\right) \approx & \ R_\ell^{AB} \, C_{\ell}^{AB}(i_z,j_z) \ R_{\ell'}^{CD} \, C_{\ell'}^{CD}(k_z,l_z) \\
& \times S^{A,B;C,D}_{i_z,j_z;k_z,l_z}, \label{Eq:approx-SSC}
\ea
where the double integrals over redshift in Eq. (\ref{Eq:exact-SSC}) have been (approximately) performed. The matrix $S^{A,B;C,D}_{i_z,j_z;k_z,l_z}$ is the dimensionless volume-averaged (co)variance of the background matter density contrast
\ba\label{Eq:def-Smatrix}
S^{A,B;C,D}_{i_z,j_z;k_z,l_z} = \int \dd V_1 \, \dd V_2 \, \frac{W_{i_z}^A(z_1) \, W_{j_z}^B(z_1)}{I^{AB}(i_z,j_z)} \, \frac{W_{k_z}^C(z_2) \, W_{l_z}^D(z_2)}{I^{CD}(k_z,l_z)} \ \sigma^2(z_1,z_2),
\ea
with
\ba\label{Eq:I-AB-vol}
I^{AB}(i_z,j_z) = \int \dd V_1 \ W_{i_z}^A(z_1) \ W_{j_z}^B(z_1).
\ea
The quantity $R_\ell$ is the effective relative response of the considered power spectrum. In the context of second order perturbation theory, the growth-only response of the matter power spectrum is $\frac{\partial P(k)}{\partial \delta_b} = \frac{68}{21} P(k)$ \citep[e.g.][]{Takada2013}, i.e. $R_\ell = \frac{68}{21}$. Other non-linear terms are however present, which increase the total response. In App. \ref{App:dClddeltab} we detail these terms, and present a full computation of the response. Our formalism is valid for a general scale-dependent response. In numerical applications later in this article, we will test the approximation both with the full response and with the simpler ansatz $R_\ell = 5\equiv R$, which is the effective value found in App. \ref{App:dClddeltab}.

The key point is that starting from the above approximation, Eq. (\ref{Eq:approx-SSC}), makes the computation of the SSC for angular power spectra to have the same numerical cost as the computation of the power spectra themselves.


\section{Application to parameter constraints}\label{Sect:applic}

In this section we examine the consequences for data analysis or forecasts of the SSC covariance given by Eq.~(\ref{Eq:approx-SSC}) as an update to the covariance, i.e. the total covariance is $\mathcal{C} = \mathcal{C}_\mr{noSSC} + \mathcal{C}_\mr{SSC}$, where we note $\mathcal{C}_\mr{noSSC}$ the sum of all other contributions to the covariance matrix.

A common statistical use of a covariance matrix, $\mathcal{C}$, is to compute scalar quantities of the form
\footnote{For all these cases of interest, the matrix-vector products in Eq. (\ref{Eq:X-Cm1-Y}) have to be understood as
\ba
	I=\displaystyle\sum_{\ell,\ell'=\ell_\mr{min}}^{\ell_\mr{max}} X_\ell~\left[\mathcal{C}^{-1}\right]_{\,\,\ell\ell'}~Y_{\ell'}, \nonumber
\ea
i.e. it correspond to a cumulative quantity over multipole.}
\ba\label{Eq:X-Cm1-Y}
I = X^T \cdot \mathcal{C}^{-1} \cdot Y,
\ea
where $\cdot$ stands for matrix multiplication. For example to compute the {\it cumulative} signal-to-noise ratio (dubbed $(S/N)$ hereafter) one would have $X=Y=\left(C_\ell\right)_{\ell=\ell_\mr{min}\cdots\ell_\mr{max}} \equiv C$. In the exponent of a Gaussian likelihood, one would need $X=Y=\hat{C}-C(\mathbf{p})$, where $\hat{C}\equiv\left(\hat{C}_\ell\right)_{\ell=\ell_\mr{min}\cdots\ell_\mr{max}}$ is the estimated/measured power spectrum and $C(\mathbf{p}) \equiv \left(C_\ell(\mathbf{p})\right)_{\ell=\ell_\mr{min}\cdots\ell_\mr{max}}$ is the predicted power spectrum with model parameters $\mathbf{p}$. Finally for Fisher forecasts, computing the Fisher matrix, $F_{\alpha,\beta}$, requires $X=\partial C / \partial \mathbf{p}_\alpha \equiv \partial_\alpha C$ and $Y=\partial_\beta C$.

This last case is the primary aim of the article since it is a measure of the amount of informations one has on cosmological parameters from the observables $C$. The Fisher matrix will be our figure of merit to gauge the quality of the approximation, with numerical results to be presented in Sect. \ref{Sect:Num-Fisher}. \\

Computing the scalar quantities given by Eq. (\ref{Eq:X-Cm1-Y}) requires the inversion of the covariance matrix, $\mathcal{C}=\mathcal{C}_\mr{noSSC}+\mathcal{C}_\mr{SSC}$. Using the approximation Eq. (\ref{Eq:approx-SSC}), adding the SSC corresponds to a rank 1 update of the covariance matrix of the angular power spectrum $C_\ell$. Furthermore, in App. \ref{App:binning}, we detail the way to introduce binned power spectra in this approach, and we show that the adding the SSC to the covariance of the {\it binned} spectra is also a rank 1 update of the covariance.\\
We will thus make use of the Sherman-Morrison formula \citep{Sherman1950,Bartlett1951} which gives the impact on matrix inversion of a rank 1 update:
\ba\label{Eq:Sherman-Morrison}
\left(A+U V^T\right)^{-1} = A^{-1} - \frac{\left(A^{-1} \cdot U\right) \left( V^T \cdot A^{-1}\right)}{1 + V^T \cdot A^{-1} \cdot U},
\ea
where $A$ is any $n\times n$ square matrix, and $U$ and $V$ are two $n$-dimensional vectors, and $T$ means the transpose.

\subsection{Single probe and single redshift bin}\label{Sect:Fisher-singlep-singleb}

If we neglect all non-Gaussian terms except SSC, the noSSC covariance reduces to the Gaussian one which is diagonal in full sky:
\ba
\left(\mathcal{C}_\mr{G}\right)_{\ell,\ell'} = G_\ell \ \delta_{\ell,\ell'} \qquad \mr{with} \qquad G_\ell = \frac{2 C_\ell^2}{2\ell+1}.
\ea
This can simplify the inversion of the noSSC covariance later on.
In partial sky observations, the Gaussian covariance will not be diagonal due to mask-induced couplings between different angular scales (i.e. $\ell$-to-$\ell'$ couplings). It can nevertheless be made diagonal in practice by binning the power spectrum with bins wider than the typical width of the mask-induced couplings, as shown in App. \ref{App:binning}.

In the following we will keep the covariance general throughout the derivation, and indicate when appropriate which expressions are simplified by the diagonal assumption. We will also keep the same subscript, $\ell$, to label either single multipoles or bins of multipoles.\\

The super-sample covariance has a separable form between the two multipoles so that we can write the total covariance as
\ba
\mathcal{C} = \mathcal{C}_\mr{noSSC} + S_{i,i} \ \left(V  V^T\right)
\ea
where $V$ is a vector with size the number of multipoles, given by
\ba
V_\ell = R_\ell \, C_{\ell},
\ea
and $S_{i,i}$ is just a number. The Sherman-Morrison formula Eq. (\ref{Eq:Sherman-Morrison}) then gives the inverse covariance as 
\ba
\mathcal{C}^{-1} = \mathcal{C}_\mr{noSSC}^{-1} - \frac{S_{i,i} \ \left(\mathcal{C}_\mr{noSSC}^{-1} V\right) \left(V^T \mathcal{C}_\mr{noSSC}^{-1}\right)}{1+S_{i,i} \ \left(V^T \cdot \mathcal{C}_\mr{noSSC}^{-1} \cdot V\right)}
\ea
where $V^T \cdot \mathcal{C}_\mr{noSSC}^{-1} \cdot V$ is a scalar.

Thus the scalar quantity defined in Eq. (\ref{Eq:X-Cm1-Y}) is given by
\ba\label{Eq:I-singleprobe-singlez}
I = I_\mr{noSSC} - \frac{f^\mr{SSC}_X \ f^\mr{SSC}_Y \ S_{i,i}}{1+V^T \cdot \mathcal{C}_\mr{noSSC}^{-1} \cdot V \ S_{i,i}},
\ea
where we defined the scalar
\ba
f^\mr{SSC}_X \equiv X^T \cdot \mathcal{C}_\mr{noSSC}^{-1} \cdot V \stackrel{\text{diagonal}}{=} \sum_\ell \frac{R_\ell \, C_{\ell} \ X_\ell}{G_\ell}.
\ea
The notation $\stackrel{\text{diagonal}}{=}$ in the above means that the assumption that the noSSC covariance matrix is diagonal has been used. In particular for Fisher matrices, SSC gives a negative correction to the noSSC case:
\ba\label{Eq:Fisher-singleprobe-singlez}
F_{\alpha,\beta} = F^\mr{noSSC}_{\alpha,\beta} - \frac{f^\mr{SSC}_\alpha \ f^\mr{SSC}_\beta \ S_{i,i}}{1+V^T \cdot \mathcal{C}_\mr{noSSC}^{-1} \cdot V \ S_{i,i}},
\ea
with
\ba
f^\mr{SSC}_\alpha \equiv \partial_\alpha C^T \cdot \mathcal{C}_\mr{noSSC}^{-1} \cdot V \stackrel{\text{diagonal}}{=} \sum_\ell \frac{R_\ell \, C_{\ell} \ \partial_\alpha C_\ell}{G_\ell}.
\ea
We finally mention that all the above expressions for the impact of the SSC are easily extended to the case of binned spectra by replacing the vector $V_\ell$ by its binned version, $V_b$ (see App. \ref{App:binning}).

\subsection{Multi-probe and single redshift bin}\label{Sect:Fisher-multip-singleb}

A more complex case of interest is when we have spectra of different probes sharing the same redshift bin. One example is a multi-tracer analysis for galaxy clustering at a given redshift. Another is a weak-lensing analysis splitting galaxy types (e.g. red vs blue galaxies) to better mitigate the effect of intrinsic alignments.

In the following we are going to illustrate with two probes ($A$ and $B$) yielding three power spectra, although the results hold straightforwardly for more probes. \\

The three angular power spectra are going to be generally correlated. So even in the Gaussian case, the covariance matrix will not be diagonal as a function of probes, e.g. even in full sky
\ba
\Cov_\mr{G}(C_\ell^{AA},C_\ell^{BB}) = \frac{2 \left(C_\ell^{AB}\right)^2}{2\ell+1} \neq 0.
\ea
However it remains diagonal as a function of multipoles.

Given that they share the same redshift bin, and assuming the weighting kernels to have similar enough redshift dependence within the bin, the $S$ matrix can be assumed independent of probes :
\ba
S^{W,X;Y,Z}_{i,i;i,i} = \mr{cst} \equiv S_{i,i} \qquad \forall \ (W,X,Y,Z)\in \{A,B\}^4.
\ea
This property is going to simplify the super-sample covariance, allowing to easily compute its impact on Fisher forecasts as we show below.

In the following we call $n_c$ the number of spectra and $n_\ell$ the number of multipoles.
It now becomes useful to arrange the $n_c\times n_\ell$ data vector $\mathbf{C}$ grouping probes together before multipoles. For example with two probes $A$ and $B$
\ba
\mathbf{C} = 
\left(\begin{matrix}
\mathbf{C}_{\ell_\mr{min}}\\
\vdots \\
\mathbf{C}_{\ell_\mr{max}}
\end{matrix}\right)
\qquad \mr{with} \qquad
\mathbf{C}_{\ell} =
\left(\begin{matrix}
C_{\ell}^{AA} \\
C_{\ell}^{AB} \\
C_{\ell}^{BB}
\end{matrix}\right)
\ea
The covariance matrix has a size $(n_c \times n_\ell)\times(n_c \times n_\ell)$. With $\mathbf{C}$ arranged as above, the covariance matrix is thus partitionned in $n_\ell \times n_\ell$ blocks, each of these blocks having a size $n_c\times n_c$.\\
In the Gaussian case, in full sky or with the $f_\mr{sky}$ approximation, the covariance matrix is block diagonal, and we call $\mathbf{G}_\ell$ these blocks of size $n_c\times n_c$ on the diagonal. (We refer to App. \ref{App:binning} for the case of binned spectra.) \\

The formalism of Sect. \ref{Sect:Fisher-singlep-singleb} for a single probe is then easily adapted with only slight changes. We have the total covariance matrix
\ba
\mathcal{C} = \mathcal{C}_\mr{noSSC} + \mathcal{C}_\mr{SSC} = \mathcal{C}_\mr{noSSC} + \mathbf{V} \mathbf{V}^T \ S_{i,i}
\ea
where $\mathbf{V}$ is a $n_c\times n_\ell$ vector, given by
\ba
\mathbf{V}_\ell = R_\ell \ \mathbf{C}_{\ell}
\ea
Then the inverse covariance follows
\ba
\nonumber \mathcal{C}^{-1} &= \mathcal{C}_\mr{noSSC}^{-1} - \frac{ S_{i,i} \ \left(\mathcal{C}_\mr{noSSC}^{-1} \mathbf{V}\right)\left( \mathbf{V}^T \mathcal{C}_\mr{noSSC}^{-1}\right)}{1+S_{i,i} \ \left(\mathbf{V}^T \cdot \mathcal{C}_\mr{noSSC}^{-1} \cdot \mathbf{V}\right)},
\ea
where $\mathbf{V}^T \cdot \mathcal{C}_\mr{noSSC}^{-1} \cdot \mathbf{V}$ is a scalar. For a block-diagonal covariance, it simplifies to
\ba
\mathbf{V}^T \cdot \mathcal{C}_\mr{noSSC}^{-1} \cdot \mathbf{V} \stackrel{\text{diagonal}}{=} \sum_\ell \mathbf{C}_\ell^T \cdot \mathbf{G}^{-1}_\ell \cdot \mathbf{C}_{\ell},
\ea
with the inner matrix products being in the space of the $n_c$ spectra, and appropriately reduce to the case of Sect. \ref{Sect:Fisher-singlep-singleb} when $n_c=1$.

Thus the scalar quantity defined in Eq. (\ref{Eq:X-Cm1-Y}) is given by
\ba\label{Eq:I-multiprobes-singlez}
I = I_\mr{noSSC} - \frac{\mathbf{f}^\mr{SSC}_X \ \mathbf{f}^\mr{SSC}_Y \ S_{i,i}}{1+\mathbf{V}^T \cdot \mathcal{C}_\mr{noSSC}^{-1} \cdot \mathbf{V} \ S_{i,i}}
\ea
where we defined the scalar
\ba
\mathbf{f}^\mr{SSC}_X \equiv \mathbf{X}^T \cdot \mathcal{C}_\mr{noSSC}^{-1} \cdot \mathbf{V} \stackrel{\text{diagonal}}{=} \sum_\ell \mathbf{X}_\ell \cdot \mathbf{G}^{-1}_\ell \cdot R_\ell \ \mathbf{C}_{\ell},
\ea
with again the inner matrix products in the space of the $n_c$ spectra.

Finally the total Fisher matrix is given by the noSSC Fisher matrix plus a negative SSC correction:
\ba\label{Eq:Fisher-multiprobes-singlez}
F_{\alpha,\beta} = F^\mr{noSSC}_{\alpha,\beta} - \frac{\mathbf{f}^\mr{SSC}_\alpha \ \mathbf{f}^\mr{SSC}_\beta \ S_{i,i}}{1+\mathbf{V}^T \cdot \mathcal{C}_\mr{noSSC}^{-1} \cdot \mathbf{V} \ S_{i,i}},
\ea
with
\ba
\mathbf{f}^\mr{SSC}_\alpha \equiv \partial_\alpha \mathbf{C}^T \cdot \mathcal{C}_\mr{noSSC}^{-1} \cdot \mathbf{V} \stackrel{\text{diagonal}}{=} \sum_\ell \partial_\alpha \mathbf{C}_\ell^T \cdot \mathbf{G}^{-1}_\ell \cdot R_\ell \ \mathbf{C}_{\ell}.
\ea

\subsection{Multi-probe and multiple redshift bins}\label{Sect:Fisher-multip-multib}

A first case of interest is when we have probes in different non-overlapping bins, or if the overlap is small enough to be neglected. This happens for instance for galaxies, cluster counts or power spectra in sufficiently wide bins, i.e. larger than the photo-$z$ error bars, and $\geq 0.1$ to be larger than the width of SSC's $\sigma^2(z_1,z_2)$ \cite[see Fig.~6 of][]{Lacasa2016}.

In such a case, we can basically add up the bins independently:
\ba
I = \sum_i I_{i,i} = I_\mr{noSSC} - \Delta I_{SSC},
\ea
where $I_{i,i}$ is given by Eq.~(\ref{Eq:I-multiprobes-singlez}). The (negative) SSC correction is
\ba
\Delta I_{SSC} = \sum_i \frac{\mathbf{f}^\mr{SSC}_X \ \mathbf{f}^\mr{SSC}_Y \ S_{i,i}}{1+\mathbf{V}^T \cdot \mathcal{C}_\mr{noSSC}^{-1} \cdot \mathbf{V} \ S_{i,i}}.
\ea
In particular for Fisher forecasts, the (negative) SSC correction reads
\ba\label{Eq:SSCcorr-Fishermultiz}
\Delta F_{\alpha,\beta}^{SSC} = \sum_i \frac{\mathbf{f}^\mr{SSC}_\alpha \ \mathbf{f}^\mr{SSC}_\beta \ S_{i,i}}{1+\mathbf{V}^T \cdot \mathcal{C}_\mr{noSSC}^{-1} \cdot \mathbf{V} \ S_{i,i}},
\ea
which is just obtained as the sum over independent redshift bins of the SSC corrections derived for one single bin. \\

The second case of interest is when bins are overlapping. This happens for instance when analyzing galaxy shear --which integrates signal from $z=0$ to the sources-- either alone or in combination with other probes. In that case, no simplifications can be carried out: the power spectra are correlated both as a function of multipoles and as a function of redshift bins. The covariance matrix must be built in full generality using Eq.~(\ref{Eq:approx-SSC}), and then inverted numerically. We note that already at the Gaussian level inversion must be carried out numerically, due to the coupling between redshift bins.

\subsection{Importance of SSC: an analytical rule of thumb}\label{Sect:impSSC}
The importance of SSC can be gauged easily in an analytical way, if we assume a single redshift bin, and further approximate the response $R_\ell$ to be independent of scale $R_\ell\equiv R$. We underline that this scale independent assumption is not a requirement for numerical application, and may be relaxed as will be done in Sect.~\ref{Sect:Num-Fisher}.\\
In this subsection, we will gauge the importance of SSC analytically first for the signal-to-noise ratio, then for Fisher constraints.

\subsubsection{Impact on the signal-to-noise ratio}\label{Sect:impSSC-S/N}
The $(S/N)$ of a set of angular power spectra, collected in a single data vector $\mathbf{C}$, is given by
\ba
\left(S/N\right)^2 = \mathbf{C} \cdot {\cal{C}}^{-1} \cdot \mathbf{C},
\ea
and we also introduce the noSSC version of it as
\ba
\left(S/N\right)^2_\mr{G} = \mathbf{C} \cdot \mathcal{C}_\mr{noSSC}^{-1} \cdot \mathbf{C}.
\ea
We remind that these are {\it cumulative} $(S/N)$'s obtained as a summation over multipoles up to a maximum value $\ell_\mr{max}$. It is then a function of the maximum multipole up to which we integrate our observables. Let us also introduce the scalar quantity 
\ba \label{Eq:y}
Y \equiv \mathbf{V}^T \cdot \mathcal{C}_\mr{noSSC}^{-1} \cdot \mathbf{V} \ S_{i,i}.
\ea
With the assumption of a single redshift bin and a scale independent response, the scalar $Y$ is shown to be proportional to the square of the noSSC signal-to-noise ratio, i.e.
\ba \label{Eq:ysn}
	Y=\left(S/N\right)^2_\mr{noSSC} \, R^2 \, S_{i,i}.
\ea
Then the total $(S/N)$ boils down to
\ba
\left(S/N\right)^2 = \left(S/N\right)^2_\mr{noSSC} \ \left( 1 - \frac{Y}{1+Y}\right) = \frac{\left(S/N\right)^2_\mr{noSSC}}{(1+Y)}. \label{Eq:SNR-Sij-withY}
\ea
It is thus obvious that the SSC decreases the signal-to-noise ratio compared to the noSSC case as $Y$ is by construction a positive number. Impact of the SSC is enhanced for higher values of $Y$, which is then an excellent indicator of the importance of it.\footnote{In the case of many {\it uncorrelated} redshift bins, Eq. (\ref{Eq:SNR-Sij-withY}) is generalized by summing over the redshift bins, i.e.
\ba \nonumber
\left(S/N\right)^2 = \displaystyle\sum_i \frac{\left(S/N\right)^2_{G,i}}{(1+Y_i)}.
\ea
}

From Eq. (\ref{Eq:ysn}), the impact of the SSC increases for higher signal to noise, as the higher $(S/N)_\mr{noSSC}$, the higher $Y$. Enlarging the set of power spectra to smaller angular scales (i.e. increasing $\ell_\mr{max}$ to higher multipoles) increases the signal-to-noise ratio, hence the impact of the SSC. By integrating to smaller scales, the entire signal-to-noise ratio will thus reach a plateau at an asymptotic value
\ba \label{Eq:snmax}
\left(S/N\right)^2_\mr{max} =\frac{(S/N)^2_\mr{G}}{Y} = \frac{1}{R^2 \, S_{i,i}}. 
\ea
This saturation is reached when $Y\sim 1$. For the case of a full sky cosmic variance-limited analysis of a single power spectrum up to a maximum multipole $\ell_\mr{max}$, and neglecting other non-Gaussian terms, we have $Y\sim \frac{\ell_\mr{max}^2}{2} R^2 S_{i,i}$. The typical angular scales above which the $(S/N)$ starts to saturate because of the SSC is defined by $Y\simeq1$. We thus find that SSC becomes important when the analysis goes up to $\ell_\mr{max}\gtrsim\ell_\mr{SSC}$ given by
\ba
\ell_\mr{SSC} = \sqrt{\frac{2}{R^2 \ S_{i,i}}}.
\ea

Generalising to the case of partial sky coverage and several cosmic variance-limited probes (in the same redshift bin and neglecting other non-Gaussian terms), the analysis will be affected as soon as it reaches multipoles of order
\ba
\ell_\mr{SSC} = \sqrt{\frac{2}{N_p^\mr{eff} \ R^2 \ f_\mr{sky} \ S_{i,i}}}.
\ea
where
\ba
N_p^\mr{eff} = \frac{2 \ (S/N)^2_\mr{G,joint}}{f_\mr{sky} \ \ell_\mr{max}^2}
\ea
is the effective number of probes\footnote{It is exactly the number of probes if they are uncorrelated, but it goes down to 1 if they are totally correlated.}.\\
Finally, if the probes are not cosmic variance-limited, e.g. due to the presence of shot-noise (galaxy clustering) or shape noise (weak lensing), or if other non-Gaussian covariance terms are important, then one needs a full computation of the noSSC signal to noise ratio accounting for these additional sources of error. And the criterion for the importance of SSC is $(S/N)_\mr{noSSC}^2 \gtrsim 1/(R^2 S_{i,i})$. We note that this latter critical value is that of the maximum signal to noise with the full covariance (Eq.(\ref{Eq:snmax})), this plateau being the same for single- and multi-probe cases (as long as all probes have the same response $R$). i.e. it is the maximum amount of information that can be extracted from matter fluctuations in a finite volume of the universe with probes with a given response, regardless of the number of probes.

\subsubsection{Impact on Fisher constraints}\label{Sect:impSSC-Fisher}

The signal-to-noise ratio is highly impacted by SSC, because in the SSC dominated regime and with a constant $R_\ell$, Eq.~(\ref{Eq:approx-SSC}) tells that $C_\ell$ measurements are 100\% correlated and thus all information is lost on the overall amplitude. However one may question the impact on cosmological parameters, if those are sensitive to other features of the power spectrum.

We first remind Eq.~(\ref{Eq:Fisher-multiprobes-singlez}) for the Fisher information on model parameters $\alpha$ and $\beta$:
\ba
F_{\alpha,\beta} = F^\mr{noSSC}_{\alpha,\beta} - \frac{\mathbf{f}^\mr{SSC}_\alpha \ \mathbf{f}^\mr{SSC}_\beta \ S_{i,i}}{1+Y},
\ea
rewritten here using $Y$. We further introduce two angles: first the angle $\theta_\alpha$ between the vectors $\mathbf{C}$ and $\partial_\alpha \mathbf{C}$
\ba
\cos\theta_\alpha =& \frac{\partial_\alpha \mathbf{C}^T \cdot \mathcal{C}_\mr{noSSC}^{-1} \cdot \mathbf{V}}{\sqrt{\partial_\alpha \mathbf{C}^T \cdot \mathcal{C}_\mr{noSSC}^{-1} \cdot \partial_\alpha \mathbf{C} \times \mathbf{V}^T \cdot \mathcal{C}_\mr{noSSC}^{-1} \cdot \mathbf{V}}} \label{Eq:def-costheta} \\
=& \frac{\mathbf{f}^\mr{SSC}_\alpha}{\sqrt{F^\mr{noSSC}_{\alpha,\alpha} \times Y/S_{i,i}}}, \nonumber
\ea
and second, the angle $\theta_{\alpha\beta}$ between $\partial_\alpha \mathbf{C}$ and $\partial_\beta \mathbf{C}$
\ba
\cos\theta_{\alpha\beta} =& \frac{\partial_\alpha \mathbf{C}^T \cdot \mathcal{C}_\mr{noSSC}^{-1} \cdot \partial_\beta \mathbf{C}}{\sqrt{\partial_\alpha \mathbf{C}^T \cdot \mathcal{C}_\mr{noSSC}^{-1} \cdot \partial_\alpha \mathbf{C} \times \partial_\beta \mathbf{C}^T \cdot \mathcal{C}_\mr{noSSC}^{-1} \cdot \partial_\beta \mathbf{C}}} \\
=& \frac{F^\mr{noSSC}_{\alpha,\beta}}{\sqrt{F^\mr{noSSC}_{\alpha,\alpha} \times F^\mr{noSSC}_{\beta,\beta}}}. \nonumber
\ea
Let us roughly interpret these angles. The second, $\theta_{\alpha\beta}$, is easily interpreted as the noSSC correlation between the parameter $\alpha$ and the parameter $\beta$. The first angle, $\theta_\alpha$, can be interpreted as follows. Reminding that $\mathbf{V}$ is proportional to the data vector $\mathbf{C}$. Up to a normalization constant, $\mathbf{C}$, hence $\mathbf{V}$, can be viewed as $\partial_A\mathbf{C}$ where $A$ is the normalization of the data vector. Since angles are obtained from normalized vectors, $\theta_\alpha$ is thus a measure of the noSSC correlation between the parameter $\alpha$ and the normalization of the data vector.

The Fisher information matrix including SSC is then conveniently expressed as a function of the noSSC Fisher matrix, the parameters $Y$ measuring the impact of the SSC on the signal-to-noise ratio, and the angles $\theta_\alpha,~\theta_\beta$, and $\theta_{\alpha\beta}$, i.e.
\ba\label{Eq:Fisher-wSSC-cos}
F_{\alpha,\beta} = F^\mr{noSSC}_{\alpha,\beta} \left(1-\frac{\cos\theta_\alpha \cos\theta_\beta}{\cos\theta_{\alpha\beta}} \frac{Y}{1+Y}\right).
\ea
The impact of the SSC on the Fisher matrix is driven first by the impact of the SSC on the $(S/N)$ through $Y$, and second, by the angles $\theta_\alpha,~\theta_\beta$ and $\theta_{\alpha\beta}$. In particular for the diagonal elements, the change of the Fisher matrix is
\ba
\frac{\delta F_{\alpha,\alpha}}{F^\mr{noSSC}_{\alpha,\alpha}} = -\cos^2\theta_\alpha \left(\frac{Y}{1+Y}\right).
\ea
This is negative-valued showing that the SSC lowers the amount of information on a given parameter. 

Two conditions have to be met for the impact of the SSC to be important: first $Y$ should be greater than one, and second $\cos^2\theta_\alpha$ should be close to one. Supposing $Y\gg1$ and if $\theta_\alpha \neq 0,\pi$, i.e. $\cos\theta_\alpha\neq\pm 1$,\footnote{The case $\cos\theta_\alpha=\pm 1$ corresponds to the parameter $\alpha$ being $\pm$ the amplitude of the power spectrum. In that case, up to a normalization we are going back to the case of the signal to noise ratio studied in Sect.~\ref{Sect:impSSC-S/N}.} then the Fisher information keeps increasing with $\ell_\mr{max}$, but at a reduced rate compared to the noSSC case, with the asymptote
\ba
F_{\alpha,\alpha} \sim F^\mr{noSSC}_{\alpha,\alpha} \left(1-\cos^2\theta_\alpha\right),
\ea
i.e. the unmarginalised error bar is increased as
\ba
\sigma_\alpha \sim \frac{\sigma^G_\alpha}{|\sin\theta_\alpha|}.
\ea
A maximal impact of the SSC is thus obtained for $\theta_\alpha$ close to zero, that is when the parameter $\alpha$ is at the noSSC level highly correlated with the normalization of the data vector.

When we have several parameters, the situation becomes more complex and cannot be  judged with only rule of thumbs. For example a parameter $\alpha$ may seem unaffected by SSC because $\cos\theta_\alpha\ll 1$, but it may be correlated (already at the noSSC level) with a parameter $\beta$ which is affected by SSC, so that $\alpha$ will be affected indirectly through marginalisation. Another possibility is that we have two parameters which are uncorrelated at the noSSC level, but through Eq.~(\ref{Eq:Fisher-wSSC-cos}) they become correlated due to SSC ; in that case a large error on one parameter would rebound on the other, which did not happen in the noSSC case.

\paragraph*{\bf{Summary--}}To decide the importance of SSC on parameter constraints, one should first compute the multipole above which one enters in the SSC dominated regime, i.e.
\ba
\ell_\mr{SSC} = \sqrt{\frac{2}{N_p \ R_\ell^2 \ f_\mr{sky} \ S_{i,i}}}.
\ea
If the analysis is restricted to scales such as $\ell \ll \ell_\mr{SSC}$, then it is not going to be affected. If the analysis enters the SSC dominated regime, then for each parameter of interest one needs to compute the angle
\ba
\cos\theta_\alpha =& \frac{\lbra \partial_\alpha \mathbf{C}, \mathbf{V}\rbra}{\| \partial_\alpha \mathbf{C}\| \ \| \mathbf{V}\|} \qquad \mr{where} \quad \lbra X, Y\rbra = X^T \cdot \mathcal{C}_\mr{G}^{-1} \cdot Y,
\ea
and the (unmarginalised) error bar on parameter $\alpha$ is going to be increased asymptotically as
\ba
\sigma_\alpha \sim \frac{\sigma^G_\alpha}{|\sin\theta_\alpha|}.
\ea
In the case with several cosmological parameters and/or nuisance parameters that need to be marginalised over, a full computation is necessary.


\section{Numerical application to Fisher forecasts}\label{Sect:Num-Fisher}

To test the accuracy of the proposed SSC approximation, as well as to illustrate its impact on a cosmological analysis, we perform here an application to a forecast of cosmological constraints from photometric galaxy clustering with the following specifications:
\begin{itemize}
\item single redshift bin with a top-hat window $0.9<z<1$, and galaxy numbers representative of Euclid : 28M galaxies in the bin, corresponding to a density $\sim 2.5$ gal/arcmin$^2$;
\item full sky coverage and an analysis in the multipole range $50<\ell<2000$, in bins of constant width $\Delta\ell=50$ (this range of scale is a realistic cut for a conservative photometric galaxy cosmological analysis relying on a constant bias model);
\item flat $w$CDM model with fiducial cosmological parameters from 
Planck 2013 $\Lambda$CDM constraints \citep{Planck2013-params}:\\
$$\{\Omega_b h^2,\Omega_{c} h^2,n_s,\sigma_8,H_0,w\}=\{0.022,0.12,0.96,0.83,67,-1\}.$$
\item we neglect other non-Gaussian covariance terms beyond SSC
\end{itemize}

With these specifications, we compute the galaxy angular power spectrum $C_\ell^\mr{gal}$ using the halo model and Halo Occupation Distribution as done in \cite{Lacasa2016}. On the one hand we find that the shot-noise level is completely negligible, so that we are signal-dominated over all the scales of interest. On the other hand, we find that the non-linear part of the power spectrum is important on those scales, with the 1-halo term dominating the 2-halo one for $\ell>800$.

For the $S$ matrix, computed following Appendix \ref{App:examples-Sij}, we find the value
\ba
S_{i,i} = 6.2 \times 10^{-7}.
\ea
Following Sect. \ref{Sect:impSSC-S/N} and assuming a scale independent response $R=5$, this translates into a knee multipole $\ell_\mr{SSC} \sim 360$, and a plateau at a signal to noise ratio $S/N\sim 250$.\\

\begin{figure}[t]
\begin{center}
\includegraphics[width=.9\linewidth]{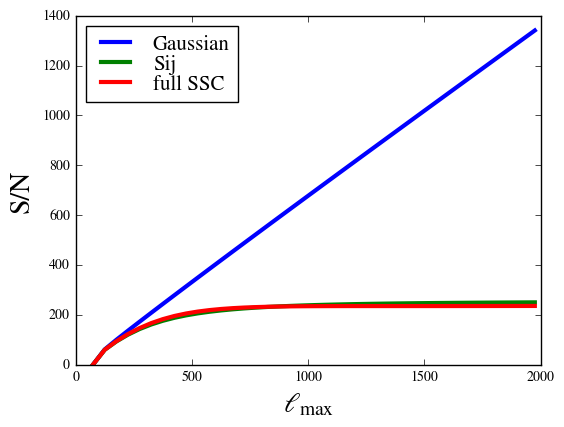}
\caption{Comparison of the cumulative signal-to-noise ratios up to a multipole $\ell_\mr{max}$ with different covariances: Gaussian only, with the $S_{i,j}$ approximation, and with a full SSC computation.}
\label{Fig:SNR-G-vs-effSij-vs-fullSSC}
\end{center}
\end{figure}

The \emph{cumulative} signal-to-noise ratio as a function of the maximum multipole of the analysis $\ell_\mr{max}$ is shown in Fig. \ref{Fig:SNR-G-vs-effSij-vs-fullSSC}. The $(S/N)$ is computed with three different covariance matrices: Gaussian only, Gaussian + SSC through Eq.~(\ref{Eq:SNR-Sij-withY}), and Gaussian plus a full SSC computation following \citet{Lacasa2016}. We see that the Gaussian covariance matrix completely overestimates the significance of the angular power spectrum, by a factor $\sim 5.7$ at $\ell_\mr{max}=2000$. However the $S_{i,j}$ approximation with a constant response does recover precisely the full SSC computation over all the multipole range, with a precision better than 7\%. We also see that the value of the knee multipole and the $(S/N)$ plateau mentioned previously indeed capture the features of the full SSC curve. The $S_{ij}$ approximation is thus validated at the level of the signal-to-noise ratio. \\

\begin{figure}[!th]
\begin{center}
\includegraphics[width=.9\linewidth]{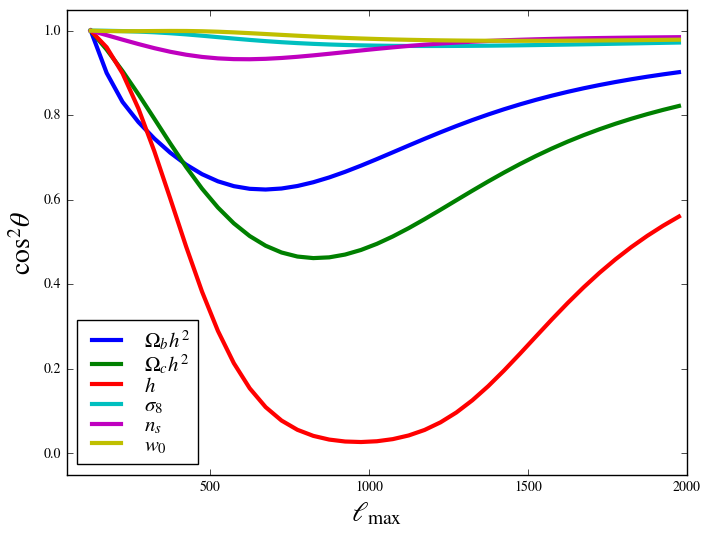}
\caption{$\cos^2\theta$ for each cosmological parameter as a function of the maximum multipole $\ell_\mr{max}$. Parameters with $\cos^2\theta$ close to 1 are going to be the most affected by SSC when it starts to dominate the covariance (i.e. for multipoles $\ell\gtrsim\ell_\mr{SSC}$ with $\ell_\mr{SSC}~360$ in this specific case).}
\label{Fig:cos2theta-vs-lmax}
\end{center}
\end{figure}

\begin{figure*}[!th]
\begin{center}
\includegraphics[width=.9\linewidth]{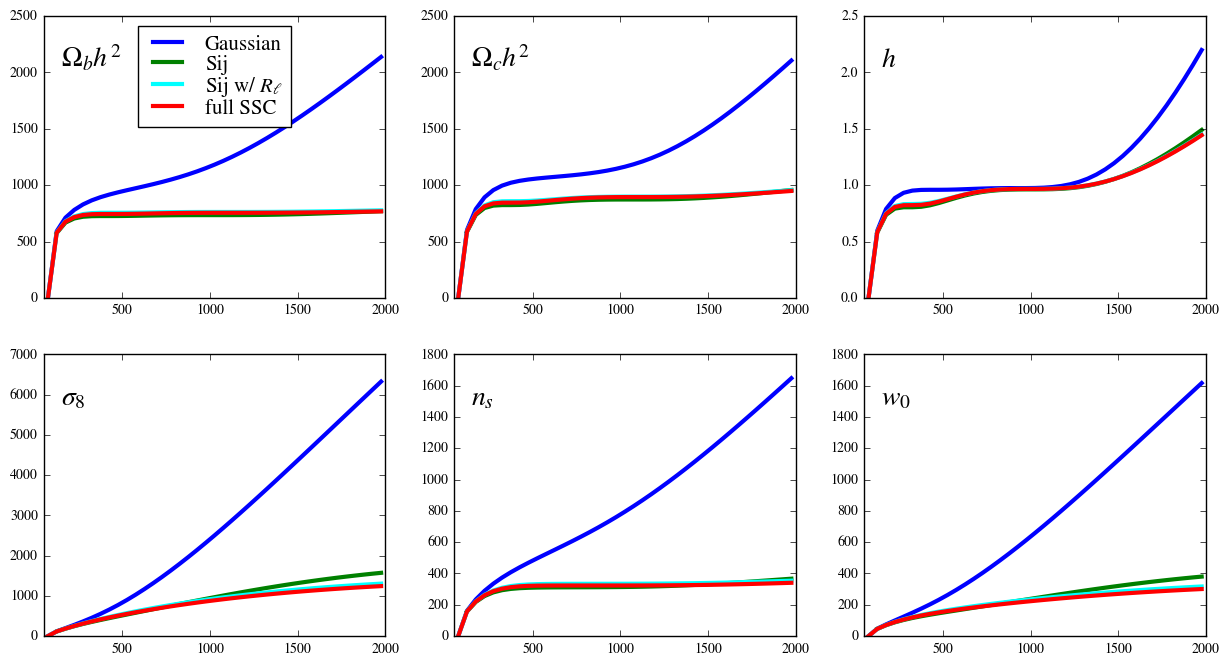}
\caption{Comparison of $\sqrt{F_{\alpha\alpha}}$ for an analysis up to a multipole $\ell_\mr{max}$ with different covariances: Gaussian only (blue), $S_{i,j}$ approximation with a constant response (green), $S_{i,j}$ approximation with a scale-dependent response (cyan), full SSC computation (red).}
\label{Fig:Fisher-G-vs-effSij-vs-fullSSC}
\end{center}
\end{figure*}

The impact of the SSC on the cosmological parameter estimation is depicted on Figs. \ref{Fig:cos2theta-vs-lmax} \& \ref{Fig:Fisher-G-vs-effSij-vs-fullSSC}. Fig. \ref{Fig:cos2theta-vs-lmax} shows the angle $\cos^2\theta_\alpha$, defined by Eq.~(\ref{Eq:def-costheta}), for the main cosmological parameters of the $w$CDM model considered here. Interestingly, all curves are significantly different from zero. Following section \ref{Sect:impSSC-Fisher}, this means that all parameters are going to be significantly impacted by SSC when reaching small scales (i.e. $\ell\gtrsim\ell_\mr{SSC}$). In this specific case, $\cos^2\theta_\alpha$ significantly different from zero will thus matters for an $\ell_\mr{max}$ greater than $\sim 400$. Some parameters should be less impacted ($\Omega_b h^2$, $\Omega_c h^2$ and $h$), while others more ($\sigma_8$, $n_s$ and $w_0$).

Fig. \ref{Fig:Fisher-G-vs-effSij-vs-fullSSC} confirms the qualitative conclusion of Fig. \ref{Fig:cos2theta-vs-lmax}. It shows for each cosmological parameter the (square root of) the Fisher element as a function of the maximum multipole $\ell_\mr{max}$ of analysis. This coefficient has an evolution with $\ell_\mr{max}$ similar to the signal-to-noise ratio,\footnote{The {\it unmarginalized} signal-to-noise ratio on a given parameter $\alpha$ is simply obtained by multiplying this coefficient with the input value of the parameter. We note also that it would coincide with the signal-to-noise ratio in the case of a model parameter being the amplitude $A$ of the power spectrum: $C_\ell = A C_\ell^\mr{template}$} and its inverse is the error bar on the considered cosmological parameter when the other parameters are fixed.\footnote{We do not attempt any marginalisation nor production of realistic forecasts. Indeed our framework is different from usual cosmological analyses, since we use the halo model and HOD, so for instance we cannot marginalise over galaxy bias.} As in the case of the $(S/N)$, on large scales ($\ell_\mr{max}<\ell_\mr{SSC}$) the impact of the SSC is negligible. This happens in spite of the angle $\cos^2\theta_\alpha$ being close to one for all the parameters, simply because the impact of the SSC on the $(S/N)$ is subdominant: $Y\ll1$. When we integrating the signal to smaller scales however ($\ell_\mr{max}\gtrsim\ell_\mr{SSC}$), the Gaussian covariance significantly overestimates the strength of the angular power spectrum: the Fisher elements become significantly smaller with the full covariance compared to the Gaussian case. Furthermore, the parameters most affected by SSC are indeed those that were identified in Fig. \ref{Fig:cos2theta-vs-lmax}: $\sigma_8$, $n_s$ and $w_0$, although the other parameters are still significantly affected. We note that this finding is in agreement with the recent analysis of \cite{Barreira2018b} in the different case of weak lensing. 

Let us briefly discuss the specific case of $h$ which is paradigmatic of how the SSC impacts on the estimation of cosmological parameters via the interplay between $Y$ and $\cos^2\theta_\alpha$. For $\ell_\mr{max}<\ell_\mr{SSC}$, the Gaussian covariance and the full covariance gives the same results since $Y\ll1$. Then for a range of multipoles roughly given by $500\lesssim\ell_\mr{max}\lesssim1500$, the Gaussian covariance and the full covariance do not give the same signal to noise ratio ; the impact of SSC on $h$ is however small in this range of multipole because the angle $\cos^2\theta_h$ is close to zero as shown in Fig. \ref{Fig:cos2theta-vs-lmax}, which suppresses the impact of the SSC. Finally, for $\ell_\mr{max}\gtrsim1500$, both $Y\gg1$ and $\cos^2\theta_h\sim1$, and the impact of the SSC is clearly seen since the Gaussian covariance now overestimates $F_{h,h}$ as compared to the full covariance. \\

Comparing the $S_{i,j}$ approximation with a constant response $R$ to the full SSC computation, we find that the former reproduces the Fisher elements of the less affected parameters ($\Omega_b h^2$, $\Omega_c h^2$ and $h$) to 3\% precision, the Fisher on $n_s$ to <8\% precision, but it is less precise for two of the heavily affected parameters ($\sigma_8$ and $w_0$) where the precision is only <30\% at the highest $\ell_\mr{max}$. This is potentially an issue for application to surveys as Euclid with requirements of 10\% precision of marginalised errors on cosmological parameters, if the analysis is pursued to these small scales. We found that the approximation respect 10\% precision on SSC up to $\ell_\mr{max}=1000$, but becomes less precise afterwards. We tracked the issue to originate from the assumption of a constant response $R_\ell=\dd\ln C_\ell/\dd\delta_b$. Using the proper scale-dependent response shown in App. \ref{App:dClddeltab}, we found that the $S_{i,j}$ approximation reproduces the Fisher elements of the full SSC computation to 5\% precision over the whole multipole range.

The $S_{i,j}$ approximation is thus validated at the level of parameter constraints. In the constant response case, it reproduces the Fisher constraints to acceptable precision, except deep in the SSC-dominated regime for the most affected parameters. Accounting for the scale-dependence of the response allows to recover all parameter constraints to sufficient precision, if one needs to pursue the analysis to small scales.

\section{Generalisations of the SSC approximation}\label{Sect:generalisations}
\subsection{Generalisation to other statistics}\label{Sect:other-stats}
A first note is that for 3D statistics, there is no need for an approximation like Eq.~(\ref{Eq:approx-SSC}). Indeed in such cases, analyses commonly use the (often implicit) assumption of no redshift evolution within the volume. And that assumption means that the SSC covariance already takes the same form as Eq.~(\ref{Eq:approx-SSC}). We refer the interested reader to e.g. \cite{Takada2013} for the 3D matter power spectrum $P(k)$.

\paragraph*{\bf Number counts--}
The case of cluster number counts is where the $S_{i,j}$ approximation in fact first started, devised by \cite{Hu2003}. Among other counts of interest for LSS surveys are those of galaxies and shear peaks. Generally we can note $N_\alpha(i_z)$ the counts with an index $\alpha$ specifying the type of object as well as the bin of the considered property (e.g. mass, luminosity, color, shear signal-to-noise, etc.). The response of such counts is the first order bias:
\ba
\frac{\partial N_\alpha}{\partial \delta_b} = b_\alpha \ N_\alpha.
\ea
i.e. $R_\alpha = b_\alpha$. The analog of Eq.~(\ref{Eq:approx-SSC}) is then
\ba
\Cov_\mr{SSC}\left(N_\alpha(i_z),N_\beta(j_z)\right) \approx & \ b_\alpha(i_z) \, N_\alpha(i_z) \ b_\beta(j_z) \, N_\beta(j_z) \times S^{\alpha;\beta}_{i_z;j_z}.
\ea
The SSC approximation is also extended to the cross-covariance with an angular power spectrum
\ba
\nonumber \Cov_\mr{SSC}\left(N_\alpha(i_z),C_{\ell'}^{CD}(k_z,l_z)\right) \approx & \ b_\alpha(i_z) \, N_\alpha(i_z) \ R_{\ell'}^{CD} \, C_{\ell'}^{CD}(k_z,l_z) \\
& \times S^{\alpha;C,D}_{i_z;k_z,l_z}.
\ea
In the case of the angular power spectrum, we could consider the response having weak scale dependence and thus approximate $R_\ell=\mr{cst}$. In the case of counts this will generally not be the case. For instance for clusters, the bias has a strong dependence on mass and will thus vary from bin to bin.

\paragraph*{\bf Correlation function--}
The 2D correlation function is a linear transform of the angular power spectrum
\ba
w(\theta) = \sum_\ell \frac{2\ell+1}{4\pi} \ C_\ell \ P_\ell(\cos\theta).
\ea
It is thus readily seen that its SSC covariance takes the form
\ba
 \Cov_\mr{SSC}\left(w^{AB}_{i_z,j_z}(\theta),w^{CD}_{k_z,l_z}(\theta')\right) \approx & \ \tilde{w}^{AB}_{i_z,j_z}(\theta) \ \tilde{w}^{CD}_{k_z,l_z}(\theta') \times S^{A,B;C,D}_{i_z,j_z;k_z,l_z}, \label{Eq:Sij-for-wtheta}
\ea
with
\ba
\tilde{w}(\theta) = \sum_\ell \frac{2\ell+1}{4\pi} \ R_\ell \ C_\ell \ P_\ell(\cos\theta) = (R * w)(\theta)
\ea
the convolution product (denoted $*$) of the original correlation function with the response $R(\theta)$. If the response $R_\ell\equiv R$ can be assumed constant as in most of this article, then $R(\theta)$ reduces to a Dirac distribution $R(\theta)= R\times\delta(\theta)$. The convolution product of $R$ with $w$ thus simplifies to a standard product, i.e. $\tilde{w}(\theta) = R \times w(\theta)$, so that Eq.~(\ref{Eq:Sij-for-wtheta}) takes the same form as Eq.~(\ref{Eq:approx-SSC}).

\paragraph*{\bf Bispectrum--}
Analogously to Eq.~(\ref{Eq:approx-SSC}), the SSC covariance for bispectra coefficients will take the form
\ba
\nonumber \Cov_\mr{SSC}\Big(b_{\ell_1\ell_2\ell_3}^{ABC}(i_z,j_z,k_z), & b_{\ell'_1\ell'_2\ell'_3}^{DEF}(l_z,m_z,n_z)\Big) \approx \ R_{\ell_1\ell_2\ell_3}^{ABC} \, b_{\ell_1\ell_2\ell_3}^{ABC}(i_z,j_z,k_z) \\
& \times R_{\ell'_1\ell'_2\ell'_3}^{DEF} \, b_{\ell'_1\ell'_2\ell'_3}^{DEF}(l_z,m_z,n_z) \ S^{A,B,C;D,E,F}_{i_z,j_z,k_z;l_z,m_z,n_z}
\ea
The 3D bispectrum SSC was studied extensively in \cite{Chan2018}, who found that the growth-only response from perturbation theory is $R_{k_1 k_2 k_3} = \frac{433}{126}$, while the total response ranges between 4 and 6. Small BAO features are visible in these responses but should we washed out in 2D projected quantities. A constant response $R=5$ may thus give an acceptable first order approximation, as we found in this article for $C_\ell$.

\subsection{Generalisation to the likelihood}\label{Sect:likelihood}

A usual assumption is that the likelihood of the observable vector $\mathcal{O}$ (e.g. the power spectrum $C=\left(C_\ell\right)_{\ell=\ell_\mr{min}\cdots\ell_\mr{max}}$ as in most of this article)
given model parameters $\mathbf{p}$, is a multivariate Gaussian
\ba
P(\mathcal{O}|\mathbf{p}) = \mathcal{N}(\bar{\mathcal{O}}(\mathbf{p}),\mathcal{C}_\mr{tot}),
\ea
where $\mathcal{N}(X,\Sigma)$ denotes the Gaussian distribution with mean $X$ and covariance $\Sigma$.\\
Given that $\mathcal{C}_\mr{tot}=\mathcal{C}_\mr{std} + \mathcal{C}_\mr{SSC}$ (e.g. for the power spectrum $\mathcal{C}_\mr{std}=\mathcal{C}_\mr{G}$) and using properties of Gaussian distributions, this can be rewritten (artificially for the moment) as the convolution
\ba
P(\mathcal{O}|\mathbf{p}) =& \mathcal{N}(\bar{\mathcal{O}}(\mathbf{p}),\mathcal{C}_\mr{std}) * \mathcal{N}(0,\mathcal{C}_\mr{SSC}) \nonumber \\
=& P_\mr{std}(\mathcal{O}|\mathbf{p}) * \mathcal{N}(0,\mathcal{C}_\mr{SSC})
\ea
where $P_\mr{std}$ is the standard (no-SSC) likelihood. The second probability distribution function (pdf) can be interpreted physically as the probability that super-survey modes induce a shift of the observable vector.

We can then reformulate the probability in a form similar to that found for cluster counts by \cite{Lima2004} (see also App. \ref{App:likelihood-clusters}):
\ba
P(\mathcal{O}|\mathbf{p}) = \int \dd(\delta \mathcal{O}) \ P_\mr{std}(\mathcal{O} | \bar{\mathcal{O}}(\mathbf{p})+\delta \mathcal{O},\mathcal{C}_\mr{G}) \ P_\mr{SSC}(\delta \mathcal{O} | 0,\mathcal{C}_\mr{SSC})
\ea
where the shift $\delta \mathcal{O}$ is a random variable with probability $P_\mr{SSC}$, i.e. it is centered on zero and has covariance matrix $\mathcal{C}_\mr{SSC}$. \\

At first order, the observable reacts to the change of background $\delta_b$ induced by long wavelength modes, through the response $\frac{\partial \mathcal{O}}{\partial \delta_b}$ (e.g. the power spectrum response discussed in Sect. \ref{Sect:fast-approx} and App. \ref{App:dClddeltab}). Noting
\ba
\bar{\mathcal{O}}(\mathbf{p},\delta_b) = \bar{\mathcal{O}}(\mathbf{p}) + \delta \mathcal{O} = \bar{\mathcal{O}}(\mathbf{p}) + \frac{\partial \mathcal{O}}{\partial \delta_b}\delta_b
\ea
the (average) observable in a part of the universe with a background change $\delta_b$, we can rewrite the likelihood as
\ba\label{Eq:likelihood-integral-SSC}
P(\mathcal{O}|\mathbf{p}) = \int \dd\delta_b \ \underbrace{P_\mr{std}(\mathcal{O} | \bar{\mathcal{O}}(\mathbf{p},\delta_b),\mathcal{C}_\mr{G}) \ P_\mr{SSC}(\delta_b|0,S)}_{\equiv P(\mathcal{O}|\mathbf{p},\delta_b)},
\ea
where we made appear the $S$ matrix defined in Eq.~(\ref{Eq:def-Smatrix}).\\
(Note that $\delta_b$ is not a simple scalar: it depends on the pair of probes and redshift bins of the observable considered.) 

Because $\delta_b$ is the density field smoothed over very large scales (the whole survey area), it is safe to assume that it has a Gaussian distribution, i.e. $P_\mr{SSC}=\mathcal{N}(0,S)$. The same however may not be true of $P_\mr{std}(\mathcal{O}|\mathbf{p})$:  the observable may indeed not have a Gaussian likelihood. For instance in the case of cluster counts studied in \cite{Lima2004}, the observable follows a Poissonian distribution if $S=0$. In the case of the angular power spectrum, which is a quadratic quantity, the observable follows a Wishart distribution in full sky \citep[e.g.][]{Hamimeche2008}, which has an important impact on inference from low multipoles / large scales. For galaxy lensing, this has been shown to be of importance by \cite{Sellentin2018b}. In the case of the bispectrum, it is also known that the likelihood should not be Gaussian \citep{Chan2017a}, although no numerical or analytical form exist for it at the moment.\\
This is where the rewriting Eq.~(\ref{Eq:likelihood-integral-SSC}) becomes useful in practice (beyond giving a nice physical interpretation) since we can now use for $P_\mr{std}(\mathcal{O} | \bar{\mathcal{O}}(\mathbf{p},\delta_b)$ a more realistic, and possibly non-gaussian, pdf.

Hence SSC can be accounted for at the likelihood level, through the hierarchical model
\ba\label{Eq:likelihood-hierarchical-SSC}
P(\mathcal{O}|\mathbf{p},\delta_b) = P_\mr{std}\left(\mathcal{O}|\bar{\mathcal{O}}(\mathbf{p},\delta_b)\right) \times \pi(\delta_b |\mathbf{p})
\ea
where $\pi$ is the prior on $\delta_b$, i.e. $\mathcal{N}(0,S)$, where the $S$ matrix depends implicitely on cosmological parameters (and potentially on other model parameters if they affect the weighting kernels). $\delta_b$ then needs to be marginalised over to get constraints on the standard model parameters.

We note that in the separate universe approach, a region with a background change in a cosmology $\mathbf{p}$ can be simulated as a region with no background change but a different cosmology with parameters $\mathbf{p'}(\mathbf{p},\delta_b)$ \citep{Wagner2015}. Thus  the likelihood Eq.~(\ref{Eq:likelihood-hierarchical-SSC}) may be implemented with only small changes to current no-SSC likelihood pipelines:
\ba\label{Eq:likelihood-separate-universe}
P(\mathcal{O}|\mathbf{p},\delta_b) = P_\mr{std}\left(\mathcal{O}|\bar{\mathcal{O}}\left(\mathbf{p'}(\mathbf{p},\delta_b)\right)\right) \times \pi(\delta_b |\mathbf{p}),
\ea
which relieves from having to model or measure the observable's response, and means that accounting for SSC is as easy as including extra nuisance parameters.

\section{Conclusion}\label{Sect:conclusions}

We presented a fast and easy approximation for the super sample covariance of 2D projected statistics, with the study mainly focused on the angular power spectrum $C_\ell$ and generalisation to other statistics given later. Besides the considered probe, this $S_{i,j}$ approximation relies on two ingredients:
\begin{itemize}
\item the $S$ matrix which is an integral of the (linear) matter power spectrum convolved with the survey window. In the flat sky limit, computable expressions are found in the literature \citep[e.g.][]{Aguena2016}, and we gave here in App.~\ref{App:examples-Sij} expressions for the full sky and partial sky cases.
\item the probe's response. We found the simple ansatz $R=5$ to perform very well for $C_\ell$'s. It is sufficient for Euclid precision requirements on parameter constraints for cosmological parameters of $w$CDM model, and up to $\ell_\mr{max}\sim 1000$. To push to smaller scales for $\sigma_8$ and $w$, it is necessary to account for the scale dependence of $R_\ell$, which we give in Appendix~\ref{App:dClddeltab} Table~\ref{Tab:fit-response-ell-z}.
\end{itemize}
Neither of these ingredients necessitate expensive computations nor physical models additional to the usual cosmological tools necessary to predict the considered probes. The $S_{i,j}$ approximation can thus readily be implemented in cosmological prediction codes.
Furthermore we showed that SSC can be included in cosmological pipelines (either for significance quantification, Fisher forecasts or MCMC parameter estimation) through a simple correction to the Gaussian covariance case, not even spoiling the speed up induced by a diagonal covariance.

The $S_{i,j}$ approximation also allows for easily identifying which cosmological parameters are going to be affected by SSC and at which level, through the fast computation of the $\cos\theta_\alpha$ coefficients, Eq. (\ref{Eq:def-costheta}) and the scalar $Y$, Eq. (\ref{Eq:y}). 

To facilitate the use of the $S_{i,j}$ approximation by the community, we release publicly a Python code that implements it, together with examples of applications, at \url{https://github.com/fabienlacasa/PySSC}.\\

In the case of photometric galaxy clustering in a redshift bin $0.9<z<1$ and with Euclid-like specifications, we found all cosmological constraints to be heavily impacted by SSC. This motivates the necessity to include this effect in forecasts and analysis pipelines for future galaxy surveys, a task now largely eased up by the $S_{i,j}$ approximation. Furthermore we showed how this approximation can be generalised beyond the angular power spectrum to other statistics such as number counts, correlation function and bispectrum, where we indicated the corresponding probe's response.\footnote{Namely the object's bias for counts, and $R=5$ for the correlation function and the bispectrum.}

Finally, we showed that the $S_{i,j}$ approximation can be generalized at the likelihood level, relieving from having to assume a Gaussian likelihood, an assumption which is incorrect in many cases as e.g. cluster counts at high masses,  $C_\ell$'s at low $\ell$ or the correlation function on large scales, and  the bispectrum. We will explore these likelihood developments in future works.

\section*{Acknowledgements}
\vspace{0.2cm}

We thank St\'{e}phane Ili\'{c} for help with the integrated Sachs-Wolfe effect. F.L. acknowledges support by the Swiss National Science Foundation.

\bibliographystyle{aa}
\bibliography{bibliography}

\appendix

\section{Example of weighting kernels}\label{App:weighting-kernels}

This section gives the weighting kernels and 3D power spectrum needed for computing the angular power spectra, Eq. (\ref{Eq:aps}), in the convention used in this article for different LSS observables.

\paragraph*{\bf {Galaxy clustering--}}
The observable is the projected galaxy number density in a redshift bin. In this case the weighting kernel is: 
\ba
W_{i_z}^\mr{GC}(z)= \frac{n_\mr{gal}^{(i_z)}(z)}{N_\mr{gal}(i_z)},
\ea
with $n_\mr{gal}(z)$ the 3D comoving galaxy density and $N_\mr{gal}(i_z)$ the 2D number of galaxies per solid angle.The 3D power spectrum is the galaxy one, i.e. $P_\mr{gg}(k)=P_\mr{gal}(k)$ which on large scales is linked to the matter power spectrum via $P_\mr{gg}(k)=b_\mr{g}^2 \ P_\mr{m}(k)$.

\paragraph*{\bf Weak lensing / shear--}
The observable is the galaxy shear averaged over a redshift bin. In this case the weighting is:
\ba
W^{\kappa_\mr{gal}}(z)= \frac{{\cal A}}{a(z) \ r(z)} \ q_{i_z}(z),
\ea
with ${\cal A}=\frac{3}{2}\Omega_m \, \left(\frac{H_0}{c}\right)^2$ and $q_{i_z}(z) = \int \dd z' \frac{n_\mr{gal}^{(i_z)}(z')}{N_\mr{gal}(i_z)} \frac{r'-r}{r'}$ the lensing efficiency \citep[e.g.][]{Kilbinger2017}. The 3D power spectrum is the matter one $P_\mr{m}(k)$.

\paragraph*{\bf CMB lensing--}
The observable is the distortion of the CMB temperature anisotropies. In this case the weighting is:
\ba
W^{\kappa_\mr{CMB}}(z)= \frac{{\cal A}}{a(z) \ r(z)} \ \frac{r_*-r}{r_*},
\ea
with $r_*$ the comoving distance to the CMB last scattering surface \citep{Lewis2006}, and the 3D power spectrum is the matter one $P_\mr{m}(k)$.

\paragraph*{\bf Integrated Sachs-Wolfe effect--}
The observable is the iSW contribution to the temperature anisotropies of the CMB. In this case the weighting kernel is \citep{Planck2015-iSW}:
\ba
W^\mr{iSW}(k,z) = - \frac{2 {\cal A}}{k^2} \left.\frac{\dd (1+z)\, G(z)}{\dd z} \middle/ \left(G(z) \ r^2(z) \frac{\dd r}{\dd z}\right) \right.,
\ea
with $G(z)$ the linear growth function. The 3D power spectrum is the matter one $P_\mr{mm}(k)$.

We note that this kernel does not depend only on redshift but also on wavenumber $k$. But since the latter dependence is factorizable, it cancels out in the $S$ matrix and thus does not impact the applicability of the approximation Eq.~(\ref{Eq:approx-SSC}).

We also note that the Limber's approximation, used throughout this article, is poorly adapted to the iSW signal because it peaks at low multipoles. \cite{Lacasa2018b} provided expressions for super-sample covariance without Limber's approximation. We leave generalisation of the present SSC approximation to this no-Limber case to future work. However, as SSC's impact peaks on small scales, we expect that its impact on iSW constraints should be small and that the present approximation should thus be good enough to gauge its level.

\section{Particular cases for $\sigma^2(z_1,z_2)$ and $S$ matrix}\label{App:examples-sigma2-and-Sij}

\subsection{$\sigma^2(z_1,z_2)$}\label{App:examples-sigma2}
In full sky we have \citep{Lacasa2016}
\ba
\sigma^2(z_1,z_2) = \frac{1}{2\pi^2}\int k^2 \, \dd k \ P_m(k|z_{12}) \ j_0(k r_1) j_0(k r_2).
\ea
In the partial sky case, $\sigma^2(z_1,z_2)$ can be expanded in spherical harmonics \citep{Lacasa2018} to get
\ba
\sigma^2(z_1,z_2) = \frac{1}{\Omega_\mr{S}^2} \sum_\ell (2\ell+1) \ C_\ell(W) \ C_\ell^\mr{m}(z_1,z_2)\,,
\ea
where $\Omega_\mr{S}=4\pi \, f_\mr{sky}$ is the solid angle covered by the survey, $C_\ell(W)$ is the angular power spectrum of the survey mask, and $C_\ell^\mr{m}$ is the angular power spectrum of matter, given by
\ba\label{Eq:Cl-matter}
C_\ell^\mr{m}(z_1,z_2) = \frac{2}{\pi} \int k^2 \, \dd k \; j_\ell(k r_1) \, j_\ell(k r_2) \; P_\mr{m}(k|z_{12})\,.
\ea

\subsection{$S$ matrix}\label{App:examples-Sij}
Assuming that super-survey modes can be described by linear theory, the matter power spectrum writes $P_\mr{m}(k|z_{12})=G(z_1) G(z_2) P(k)$, where $G(z)$ is the growth function and we noted simply $P(k)$ the power spectrum at $z=0$.

It then results from its definition Eq.~\ref{Eq:def-Smatrix} that the $S$ matrix is given by
\ba\label{Eq:Smatrix-as-int-Pk}
S^{A,B;C,D}_{i_z,j_z;k_z,l_z} = \frac{1}{2\pi^2}\int k^2 \, \dd k \ P(k) \ \mathcal{I}^{AB}_{i_z,j_z}(k) \ \mathcal{I}^{CD}_{k_z,l_z}(k),
\ea
where
\ba\label{Eq:I-AB-bessel}
\mathcal{I}^{AB}_{i_z,j_z}(k) = \int \dd V \frac{W_{i_z}^A(z) \, W_{j_z}^B(z)}{I^{AB}(i_z,j_z)} G(z) \ \mr{Win}(k r),
\ea
where we recall
\ba \label{Eq:I-AB-vol-app}
I^{AB}(i_z,j_z) = \int \dd V \ W_{i_z}^A(z) \ W_{j_z}^B(z).
\ea
and the angle-averaged survey window is
\ba
\mr{Win}(k r) = 
\left\{
\begin{array}{ll}
j_0(kr) & \mr{full\ sky} \\
\frac{4\pi}{\Omega_S^2} \sum_\ell (2\ell+1) \ C_\ell(W) \ j_\ell(k r) & \mr{partial\ sky}
\end{array}
\right.
\ea
with $C_\ell(W)$ the power spectrum of the survey mask \citep{Lacasa2018}.\\

There is a special case where the $S$ matrix can be further simplified analytically, by assuming a full sky survey where the weighting kernel is constant within the redshift bins, and approximating the growth function at the center of the redshift bin.
Indeed if the weighting kernel is constant, we have:
\ba
W_i(z) = \frac{3}{r_\mr{max}^3(i) - r_\mr{min}^3(i)} \ \mathds{1}_{z\in[z_\mr{min}(i),z_\mr{max}(i)]}.
\ea
This happens for instance in the case of galaxy clustering with perfect redshift determinations and if the galaxy comoving density can be considered constant $n_\mr{gal}(z)=\mr{cst}$. Then Eq.~(\ref{Eq:I-AB-vol-app}) simplifies to
\ba
I^{AB}(i_z,j_z) = \frac{3 \ \delta_{i_z,j_z}}{r_\mr{max}^3(i_z) - r_\mr{min}^3(i_z)}.
\ea
This gives for Eq.~(\ref{Eq:I-AB-bessel})
\ba
\nonumber \mathcal{I}^{AB}_{i_z,j_z}(k) &\approx G(z_\mr{mean}(i_z)) \frac{3 \ \delta_{i_z,j_z}}{r_\mr{max}^3(i_z) - r_\mr{min}^3(i_z)} \int_{r_\mr{min}(i_z)}^{r_\mr{max}(i_z)} r^2\ \dd r \ j_0(kr) \\
&= \frac{3 \ G(z_\mr{mean}) \ \delta_{i_z,j_z}}{k\left(r_\mr{max}^3 - r_\mr{min}^3\right)} \left[r_\mr{max}^2 \, j_1(k r_\mr{max}) - r_\mr{min}^2 \, j_1(k r_\mr{min})\right],
\ea
which can be fed into Eq.~(\ref{Eq:Smatrix-as-int-Pk}) for the $S$ matrix, leading to
\ba
\nonumber S^{A,B;C,D}_{i_z,j_z;k_z,l_z} &= \delta_{i_z,j_z} \ \delta_{k_z,l_z} \ \frac{3 \ G\left(z_\mr{mean}(i_z)\right)}{r_\mr{max}^3(i_z) - r_\mr{min}^3(i_z)} \ \frac{3 \ G\left(z_\mr{mean}(k_z)\right)}{r_\mr{max}^3(k_z) - r_\mr{min}^3(k_z)} \\
\nonumber & \times \frac{1}{2\pi^2}\int k^2 \, \dd k \ P(k) \\
\nonumber & \left[r_\mr{max}^2(i_z) \, j_1\left(k r_\mr{max}(i_z)\right) - r_\mr{min}^2(i_z) \, j_1\left(k r_\mr{min}(i_z)\right)\right]/k \\
& \left[r_\mr{max}^2(k_z) \, j_1\left(k r_\mr{max}(k_z)\right) - r_\mr{min}^2(k_z) \, j_1\left(k r_\mr{min}(k_z)\right)\right]/k. \label{Eq:Smatrix-fullsky}
\ea
We note that this expression has become independent of the considered probes ($A,B,C,D$). Furthermore, the Bessel function $j_1$ is a sum of sines and cosines : $j_1(x)=\frac{\sin x}{x^2} - \frac{\cos x}{x}$ and thus the $S$ matrix can be expressed in terms of Fourier transforms of the matter power spectrum. Specifically defining
\ba
I_{c,n}^{\pm}(r_1,r_2) \equiv & \int \dd k \ P(k)/k^n \ \cos(k(r_1\pm r_2)), \\
I_{s,n}^{\pm}(r_1,r_2) \equiv & \int \dd k \ P(k)/k^n \ \sin(k(r_1\pm r_2)), \\
\nonumber F(r_1,r_2) \equiv & -I_{c,4}^{+}(r_1,r_2)+I_{c,4}^{-}(r_1,r_2)-(r_1+r_2)I_{s,3}^{+}(r_1,r_2) \\
& +(r_1-r_2) I_{s,3}^{-}(r_1,r_2) + r_1 r_2 \left[I_{c,2}^{+}(r_1,r_2)+I_{c,2}^{-}(r_1,r_2)\right],
\ea
and shortening
\ba
r_{-,i} \equiv r_\mr{min}(i_z), \quad r_{-,k} \equiv r_\mr{min}(k_z), \quad r_{+,i} \equiv r_\mr{max}(i_z), \quad r_{+,k} \equiv r_\mr{max}(k_z),
\ea
we have the long expression
\ba
\nonumber S_{i_z,j_z;k_z,l_z} & = \delta_{i_z,j_z} \ \delta_{k_z,l_z} \ \frac{3 \ G\left(z_\mr{mean}(i_z)\right)}{r_{+,i}^3 - r_{-,i}^3} \ \frac{3 \ G\left(z_\mr{mean}(k_z)\right)}{r_{+,k}^3 - r_{-,k}^3} \times \frac{1}{4\pi^2} \\
& \times \left[ F(r_{+,i},r_{+,k})-F(r_{+,i},r_{-,k})-F(r_{-,i},r_{+,k})+F(r_{-,i},r_{-,k}) \right]. \label{Eq:Smatrix-fullsky-FFT}
\ea
Formally, the $I_{c/s,n}^{\pm}$ are IR divergent integrals as when $k\rightarrow 0$ we have $P(k)\propto k^{n_s}$ with $n_s\sim 1$. However for every $I_{c/s,n}^{+}$ there is an opposite $I_{c/s,n}^{-}$ which carries the same divergence which is thus cancelled. Hence when applying a lower cut-off $k_\mr{min}$ to all integrals, the full expression Eq.~(\ref{Eq:Smatrix-fullsky-FFT}) is convergent when $k_\mr{min}\rightarrow 0$. Numerically, one needs to apply such cut-off and not put it too low, in order to avoid cases of large cancellations between large numbers, where numerical errors could spoil the result. Inspecting Eq.~(\ref{Eq:Smatrix-fullsky}) and recalling $j_1(x)\propto x$ when $x\rightarrow 0$, one sees that the integrand is $\propto k^2 P(k) \propto k^{2+n_s}$ in the IR, thus the integral is quickly converging. A conservative choice is thus to ensure that the start of the integral is at least one decade before the matter-radiation equality, and that the Bessel functions are in the small $x$ regime. Hence we take $k_\mr{min}=\mr{min}\{k_\mr{eq},1/r_\mr{max}\}/10$. Numerically, an upper cut-off $k_\mr{max}$ also needs to be taken. The integrals are convergent in this limit so this is a less pressing issue. With the same type of arguments as for $k_\mr{min}$, one can see that Eq.~(\ref{Eq:Smatrix-fullsky-FFT}) will be well converged if we take the value $k_\mr{max}=10\times\mr{max}\{k_\mr{eq},1/r_\mr{min}\}$

\section{Angular power spectrum response}\label{App:dClddeltab}

\cite{Takada2013} showed that the 3D matter power spectrum reacts to a change of background through two separate effects: a term from second order perturbation theory (2PT hereafter) that dominates on large scales, and a term from the 1-halo part of the spectrum called halo sample variance that dominates on small scales. For the galaxy power spectrum, it was showed that the reaction also contains terms from second order galaxy bias and shot-noise and that the contribution from second order nonlocal bias vanishes \citep{Lacasa2016,Lacasa2018b}:
\ba
\nonumber \frac{\partial P_\mr{gal}(k | z)}{\partial \delta_b} &\equiv \left(\frac{68}{21} \, b_1^\mr{gal}(k,z)^2 + 2 b_2^\mr{gal}(k,z) \, b_1^\mr{gal}(k,z)\right) \, P_\mr{m}(k|z) \\
&+ \frac{\partial P_\mr{1h}(k | z)}{\partial \delta_b}  + b_1^\mr{gal}(k=0,z)/\nbargal(z), \label{Eq:dPgalddeltab}
\ea
where in the halo model
\ba
b_i^\mr{gal}(k,z) = \int \dd M \ \frac{\dd n_h}{\dd M} \ \frac{\lbra N_\mr{gal} \rbra}{\nbargal(z)} \ u(k|M,z) \ b_i(M,z) 
\ea
and
\ba
\frac{\partial P_\mr{1h}(k | z)}{\partial \delta_b} = \int \dd M \ \frac{\dd n_h}{\dd M} \ \frac{\lbra N_\mr{gal}(N_\mr{gal}-1) \rbra}{\nbargal(z)^2} \ u(k|M,z)^2 \ b_1(M,z).
\ea
with $\frac{\dd n_h}{\dd M}$ the halo mass function, $b_i(M,z)$ the i-th order halo bias, $u(k|M,z)$ the halo profile, and $N_\mr{gal}$ given by the Halo Occupation Distribution.\\
We will call the four terms in Eq.~(\ref{Eq:dPgalddeltab}) respectively 2PT, b2, 1h and shot. The reaction of the \emph{angular} power spectrum then follows
\ba
\frac{\dd C_\ell^{gg}(i_z,j_z)}{\dd \delta_b} = \int \dd V \ \frac{n_\mr{gal}^{(i_z)}(z)}{N_\mr{gal}(i_z)} \, \frac{n_\mr{gal}^{(j_z)}(z)}{N_\mr{gal}(j_z)} \ \frac{\dd P_\mr{gal}(k_\ell|z)}{\dd \delta_b}, \label{Eq:dClgalddeltab}
\ea
and it defines the (relative) response through
\ba
\frac{\dd C_\ell^{gg}(i_z,j_z)}{\dd \delta_b} = R_\ell^{gg} \ C_\ell^{gg}(i_z,j_z). \label{Eq:resp-Clgal}
\ea
\\

\begin{figure}[b]
\begin{center}
\includegraphics[width=.9\linewidth]{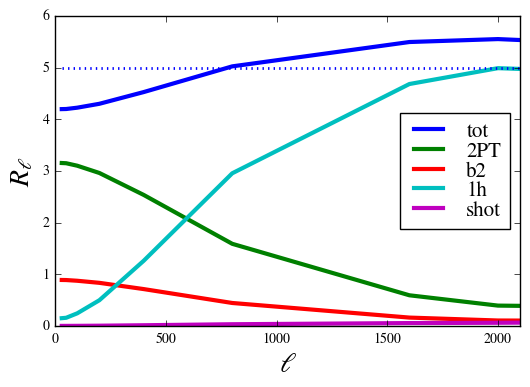}
\caption{Power spectrum response $R_\ell$ and its different terms. The dashed line indicates the effective value taken in the analysis in the main text.}
\label{Fig:response-terms}
\end{center}
\end{figure}

Fig.~\ref{Fig:response-terms} shows this response and its different terms. On large scales the response is dominated by the 2PT and b2 terms, but quickly the 1h terms start to dominate. This switch between 2h and 1h terms appears at $\ell\sim 650$, i.e. earlier than the switch in $C_\ell$ which appears at $\ell\sim 800$. This happens because the non-linear part of the power spectrum reacts more strongly to background change than the linear one : $\frac{\partial \ln P_\mr{1h}}{\partial \delta_b} > \frac{\partial \ln P_\mr{2h}}{\partial \delta_b}$.

The total angular response shows some scale dependence over the range of multipoles considered, ranging from $R_\ell\sim 4.2$ on large scales to $R_\ell\sim 5.5$ on small scales. In this article we choose for simplicity to take a constant effective value $R_\ell=5$, indicated by a dashed line in Fig.~\ref{Fig:response-terms}. This has the advantage of allowing analytical calculations in Sect.~\ref{Sect:impSSC}, and we find in Sect.~\ref{Sect:Num-Fisher} that it reproduces adequately the signal to noise ratio over all scales considered, as well as the Fisher constraints on cosmological parameters, except deep in the SSC-dominated regime for the most affected parameters.

To go beyond this constant $R_\ell$ approximation, one needs the scale dependence of the response. The redshift dependence is furthermore needed, if one wants to work on redshifts other than the one studied in this article ($0.9<z<1$). To answer both these problems, we have computed numerically the full response through Eqs.~(\ref{Eq:dPgalddeltab}) \& (\ref{Eq:resp-Clgal} on scales $50<\ell<2000$ on a wide range of redshift ($0.1<z<2$ in bins $\Delta z=0.1$). We then fitted the responses in each redshift bin either with a constant model $R_\ell=\overline{R}$ or a linear model $R_\ell=R_0+R_1\times(\ell/\ell_0)$ with $\ell_0=1000$. The values of the fitted parameters are given in Table~\ref{Tab:fit-response-ell-z}.

\begin{table}	
	\begin{center}
		\begin{tabular}{c|c|c|c}
			z & $\overline{R}$ & $R_0$ & $R_1$ \\
			\hline
			0.1 & 4.06 & 5.00 & -0.97 \\
			0.2 & 4.48 & 5.29 & -0.83 \\
			0.3 & 4.79 & 5.24 & -0.47 \\
			0.4 & 4.98 & 5.10 & -0.13 \\
			0.5 & 5.08 & 4.95 & 0.14 \\
			0.6 & 5.12 & 4.78 & 0.34 \\
			0.7 & 5.10 & 4.62 & 0.50 \\
			0.8 & 5.06 & 4.45 & 0.63 \\
			0.9 & 4.98 & 4.29 & 0.72 \\
			1.0 & 4.89 & 4.13 & 0.78 \\
			1.1 & 4.79 & 4.00 & 0.82 \\
			1.2 & 4.70 & 3.91 & 0.82 \\
			1.3 & 4.63 & 3.86 & 0.80 \\
			1.4 & 4.59 & 3.85 & 0.77 \\
			1.5 & 4.61 & 3.89 & 0.75 \\
			1.6 & 4.70 & 3.97 & 0.76 \\
			1.7 & 4.87 & 4.09 & 0.80 \\
			1.8 & 5.13 & 4.26 & 0.90 \\
			1.9 & 5.49 & 4.46 & 1.06
		\end{tabular}
	\end{center}
	\caption{Fits to the redshift dependence of the response of the galaxy power spectrum. The first column gives the start of the redshift bin, of width $\Delta z=0.1$. The second columns gives values for the scale independent model $R_\ell=\overline{R}$, while the third and fourth columns gives values for the linear model $R_\ell=R_0+R_1\times(\ell/\ell_0)$ with $\ell_0=1000$.}
	\label{Tab:fit-response-ell-z}
\end{table}

Another numerical approach that we anticipate is to calibrate the response through dedicated simulations, similarly to the work of \cite{Barreira2018}.

\section{Impact of binning}\label{App:binning}

Labelling bins of multipoles by the $b$ subscript, the binned angular power spectrum, $C_b$, is defined as 
\ba
	C_b=\displaystyle\sum_{\ell\in b}\frac{S_{b\ell}}{\Delta_b}\times C_\ell,
\ea
where the summation is over multipoles within the bin $b$, $\Delta_b$ is the width of the bin, and $S_{b\ell}$ is a reshaping operator usually chosen to flatten the $C_\ell$ within bins of multipoles.\footnote{This is easily related to the binning operator, $P_{b\ell}$, commonly used in the CMB context through $P_{b\ell}=S_{b\ell}/\Delta_b$.} This reshaping operator is thus obtained assuming that within the bin $b$, the angular power spectrum is approximately given by $C_\ell\simeq \frac{C_b}{S_{b\ell}}$ with $C_b$ a constant over the bin, and $S_{b\ell}$ a (usually theoretically) known function of $\ell$.\footnote{It is typically $\ell(\ell+1)$ in the CMB context} Finally, the width of the bins are usually chosen to be greater than the typical length in multipoles of the $\ell$-to-$\ell'$ coupling induced by the mask. \\

The covariance of the binned spectrum, $C_b$, is related to the covariance of the full spectrum, $C_\ell$, as follows
\ba
	\mathrm{Cov}\left(C_b,C_{b'}\right)=\displaystyle\sum_{\ell\in b}\sum_{\ell'\in b'}\left(\frac{S_{b\ell}}{\Delta_b}\right)\left(\frac{S_{b'\ell'}}{\Delta_b'}\right)\ \mathrm{Cov}\left(C_\ell,C_{\ell'}\right).
\ea
By writing $\mathrm{Cov}\left(C_\ell,C_{\ell'}\right)=\mathcal{C}_\mr{G}+\mathcal{C}_\mr{SSC}$, the covariance of the binned spectra is then given by the sum of its Gaussian contribution and its super-sample contribution.

 Choosing bins which are wider than the typical width of the mask-induced couplings leads to a diagonal Gaussian covariance $(\mathcal{C}_\mr{G})_{b,b'}\simeq G_b\ \delta_{b,b'}$. An analytic expression for $G_b$ can be obtained assuming the  $f_\mr{sky}$ approximation, i.e.
\ba \label{Eq:bincov}
	G_b = \displaystyle\sum_{\ell\in b}\left(\frac{S_{b\ell}}{\Delta_b}\right)^2\times\left(\frac{2 C_\ell^2}{(2\ell+1)\,f_\mr{sky} }\right).
\ea 
Since the reshaping function is chosen such as $S_{b\ell}C_\ell$ is roughly constant, one can simplify the above to get  $G_b=\frac{2C^2_{\ell_b}}{(2\ell_b+1)\,f_\mr{sky}\,\Delta_b}$ by defining the average multipole in the bin $\ell_b$ with the identification $\frac{1}{(2\ell_b+1)\Delta_b}\equiv\sum_{\ell\in b}\frac{1}{(2\ell+1)\Delta^2_b}$.

For the SSC, one first remind that for a single probe and a single redshift bin, the covariance of the spectra is given by
\ba
	\left[\mathcal{C}_\mr{SSC}\right]_{\ell,\ell'}=S_{i,i}V_\ell V_{\ell'},
\ea
with $V_\ell=R_\ell C_\ell$. It is then straighforward to show that for the binned spectra, one gets
\ba
	\left[\mathcal{C}_\mr{SSC}\right]_{b,b'}=S_{i,i}V_b V_{b'},
\ea
with the binned version of the vector $V_\ell$, i.e.
\ba
	V_b=\displaystyle\sum_{\ell\in b}\frac{S_{b\ell}}{\Delta_b}\times V_\ell.
\ea
In the case where the response $R_\ell\equiv R$ is constant, this simplifies to $V_b=RC_b$. This shows that for a single probe and a single bin in redshift, adding the SSC to covariance of {\it binned} power spectra still corresponds to a rank 1 update of the Gaussian covariance. \\

The above is easily generalized to the other cases where it is enlarged to multi-probes and more than one redshift bins since it is exactly the same binning in multipoles which has to be used for the entire set of multi-probe and multi-redshift  auto- and cross-spectra. The data vector is now built from the multi-probe {\it binned} angular power spectra, and the vector $\mathbf{V}_b$ in the SSC is obtained by binning the vector $\mathbf{V}_\ell$. Only the Gaussian covariance is slightly amended being partitioned into $n_b\times n_b$ non-diagonal blocks of size $n_c\times n_c$ ($n_b$ is the number of bins and $n_c$ the number of auto- and cross-spectra). Using the $f_\mr{sky}$ approximation, it becomes block diagonal $\mathcal{C}_\mr{G}=\mathbf{G}_b \ \delta_{b,b'}$. The blocks of size $n_c \times n_c$ reads
\ba
\left[\mathbf{G}_b\right]^{W,X;YZ} &\equiv\mathrm{Cov}_\mr{G}\left(C^{WX}_b,C^{YZ}_b\right) \nonumber\\
&= \displaystyle\sum_{\ell\in b}\left(\frac{S_{b\ell}}{\Delta_b}\right)^2\times\left(\frac{ C^{WY}_\ell C^{XZ}_\ell+C^{WZ}_\ell C^{XY}_\ell}{(2\ell+1)\,f_\mr{sky} }\right),
\ea 
where $(W,X,Y,Z)$ run over probes.

\section{Likelihood of cluster counts}\label{App:likelihood-clusters}

The purpose of this section is to recall the form of the full likelihood of cluster counts, a result which seems overlooked in the literature. Further, we will extend the likelihood with the formulation developed in Sect. \ref{Sect:likelihood}, which will ensure it to be well-defined analytically. \\

Call $N = (N_{i_M,i_z})_{i_M,i_z}$ the vector of cluster counts in all bins of mass (indexed by $i_M$) and redshift (indexed by $i_z$). Then without super-sample covariance (simply called sample variance in the cluster literature), the likelihood is a collection of independent Poisson distribution in each bin of mass and redshift:
\ba
P(N|\mathbf{p}) = \prod_{i_M,i_z} \mr{Poiss}(N_{i_M,i_z} | \overline{N}_{i_M,i_z}(\mathbf{p})),
\ea
where $\overline{N}_{i_M,i_z}(\mathbf{p})$ is the model prediction for parameters $\mathbf{p}$.

The above likelihood is sufficient to describe small counts, i.e. at high mass. However for current and future surveys detecting more and more clusters, it becomes necessary to account for the effect of sample variance \citep{Hu2003}. \cite{Lima2004} found the full likelihood for cluster counts in different cells, which can be straightforwardly applied to our case with only one cell (the survey):
\ba
P(N|\mathbf{p}) = \int \dd^n\tilde{N} \ \left(\prod_{i_M,i_z} \mr{Poiss}(N_{i_M,i_z} | \tilde{N}_{i_M,i_z}) \right) \ \mr{Gauss}(\tilde{N}-\overline{N},S'),
\ea
where $S'$ is similar to the $S$ matrix defined in Eq.~(\ref{Eq:def-Smatrix}) in the case of cluster counts, but also including the counts response and was defined originally for a 3D survey neglecting redshift evolution \citep{Lima2004}:
\ba
S'_{i_M,i_z;j_M,j_z} = b_{i_M,i_z} \, N_{i_M,i_z} \ b_{j_M,j_z} \, N_{j_M,j_z} \times S_{i_z;j_z}.
\ea
The matrix $S_{i_z;j_z}$ reads
\ba
S_{i_z;j_z} = \int \frac{\dd^3\kk}{(2\pi)^3} \ \tilde{W}_{i_z}^*(\kk) \, \tilde{W}_{j_z}(\kk) \ P(k)
\ea
where $\tilde{W}_{i_z}$ is the normalised ($\int \dd^3\xx \ \tilde{W}_{i_z}(\xx)=1$) window function in redshift bin $i_z$. \\

In the framework developed in Sect.~\ref{Sect:likelihood}, $\tilde{N}$ is interpreted as the average number counts in a region of the universe having a background change $\delta_b$. Noting that the response of the cluster counts is $\frac{\partial N_{i_M,i_z}}{\partial \delta_b} = b_{i_M,i_z} \, N_{i_M,i_z} $, we can rewrite the likelihood as
\ba
P(N|\mathbf{p}) =& \int \dd^{n}\delta N \ \left(\prod_{i_M,i_z} \mr{Poiss}\left(N_{i_M,i_z} | \overline{N}_{i_M,i_z} + \delta N_{i_M,i_z}\right) \right) \nonumber \\
& \qquad \times \mr{Gauss}\left(\delta N,\frac{\partial N}{\partial \delta_b}^T S \frac{\partial N}{\partial \delta_b}\right) \\
=& \int \dd^{n_z}\delta_b \ \left(\prod_{i_M,i_z} \mr{Poiss}\left(N_{i_M,i_z} | \overline{N}_{i_M,i_z} + \frac{\partial N_{i_M,i_z}}{\partial \delta_b}\delta_b(i_z)\right) \right) \nonumber \\
& \qquad \times \mr{Gauss}\left(\delta_b, S\right).
\ea
Rigorously, one could be concerned about the edges of the integral in this likelihood: it does not make physical sense for $\delta_b$ to go to $-\infty$, as it corresponds to $\delta N \rightarrow-\infty$, i.e. $\tilde{N}$ becoming negative which is impossible for a number of objects. In practice, this is unlikely to be a concern since for any reasonably-sized survey the background change follows $\delta_b\ll 1$ at all redshifts, hence $\tilde{N}>0$.

For the purpose of rigorousness, let us solve this physical concern nonetheless. When $\delta_b$ becomes of order 1, two approximations fail down: (i) the pdf of $\delta_b$ being Gaussian which is incorrect since for instance $\delta_b \geq -1$, and (ii) using a linear response ansatz $\tilde{N}=\overline{N}+\frac{\partial N}{\partial \delta_b} \delta_b$. Both failures can be cured formally with
\ba
P(N|\mathbf{p})=& \int \dd^{n_z}\delta_b \ \left(\prod_{i_M,i_z} \mr{Poiss}\left(N_{i_M,i_z} | \tilde{N}_{i_M,i_z}(\mathbf{p},\delta_b)\right) \right) \times P\left(\delta_b|\mathbf{p}\right),
\ea
where $P\left(\delta_b|\mathbf{p}\right)$ is the pdf of the background change (which has support on $\delta_b\in[-1,\infty[$) and $\tilde{N}_{i_M,i_z}(\mathbf{p},\delta_b)$ is the average cluster count in a region with background change $\delta_b$ of a universe with cosmological parameters $\mathbf{p}$. For instance in the separate universe approach this $\tilde{N}_{i_M,i_z}(\mathbf{p},\delta_b)$ could be computed thanks to a change of cosmological parameters $\mathbf{p'}(\mathbf{p},\delta_b)$.

\end{document}